\documentclass[11pt]{article}
\usepackage{graphicx}
\usepackage{epsfig}
\usepackage{epstopdf}
\usepackage{amsmath}
\usepackage{amsfonts}
\usepackage{amssymb}

\usepackage[english]{babel}

\newcommand{\eq}{\begin{equation}}
\newcommand{\beq}{\begin{equation}}
\newcommand{\eeq}{\end{equation}}
\newcommand{\eqa}{\begin{eqnarray}}
\newcommand{\eeqa}{\end{eqnarray}}
\newcommand{\beqa}{\begin{eqnarray}}
\newcommand{\bea}{\begin{eqnarray}}
\newcommand{\eea}{\end{eqnarray}}

\newcommand{\mc}[1]{\mathcal{#1}}
\newcommand{\dd}{{\textrm{d}}}

\newcommand{\R}{{\mathbb R}}
\newcommand{\calS}{{\mathcal S}}

\newcommand{\p}{{\partial}}
\newcommand{\al}{{\alpha}}
\newcommand{\be}{{\beta}}
\newcommand{\de}{{\delta}}
\newcommand{\ep}{{\epsilon}}
\newcommand{\vep}{{\varepsilon}}
\newcommand{\ga}{{\gamma}}
\newcommand{\ka}{{\kappa}}
\newcommand{\la}{{\lambda}}
\newcommand{\si}{{\sigma}}

\newcommand{\om}{{\omega}}
\newcommand{\Om}{{\Omega}}
\newcommand{\munu}{{\mu\nu}}

\newcommand{\lp}{\left(}
\newcommand{\rp}{\right)}

\newcommand{\lb}{\left[}
\newcommand{\rb}{\right]}

\newcommand{\eg}{{\it e.g.,}\ }
\newcommand{\ie}{{\it i.e.,}\ }

\newcommand{\lsim}{\mathrel{\raisebox{-.6ex}{$\stackrel{\textstyle<}{\sim}$}}}

\numberwithin{equation}{section}
\begin{document}
\setlength{\unitlength}{1mm}

\thispagestyle{empty}
\begin{flushright}
\small \tt
\begin{tabular}{l}
CPHT-RR111.1210\\
LPT-ORSAY 10-106
%\\\today
\end{tabular}
\end{flushright}
%\rightline{\small hep-th/08}
\vspace*{1.cm}

\begin{center}
{\bf \LARGE Higher-dimensional Rotating Charged Black Holes}\\

\vspace*{1.5cm}

{\bf Marco M.~Caldarelli,}$^{1,2,3}\,$
{\bf Roberto Emparan,}$^{4,5}\,$ {\bf Bert Van Pol,}$^3\,$

\vspace*{0.5cm}

{\it $^1$\,Laboratoire de Physique Th\'eorique, Univ. Paris-Sud,\\
 CNRS UMR 8627,
 F-91405 Orsay, France}\\[.3em]

{\it $^2$\,Centre de Physique Th\'eorique, Ecole Polytechnique,\\
 CNRS UMR 7644,
 F-91128 Palaiseau, France}\\[.3em]

{\it $^3\,$Instituut voor Theoretische Fysica, Katholieke Universiteit Leuven, \\
Celestijnenlaan 200D B-3001 Leuven, Belgium}\\[.3em]

{\it $^4\,$Instituci\'o Catalana de Recerca i Estudis Avan\c{c}ats (ICREA),
\\
Passeig Llu\'{\i}s Companys 23, E-08010 Barcelona, Spain}\\[.3em]
{\it $^5\,$Departament de F{\'\i}sica Fonamental and
Institut de Ci\`encies del Cosmos, \\
Universitat de
Barcelona,
Marti i Franqu{\`e}s 1,
E-08028 Barcelona}\\[.3em]

\vspace*{0.3cm} {\small\tt
marco.caldarelli@th.u-psud.fr,
emparan@ub.edu,
bert.vanpol@student.kuleuven.be}

\vspace*{.3cm}

\vspace{.8cm} {\bf ABSTRACT}
\end{center}

Using the blackfold approach, we study new classes of higher-dimensional
rotating black holes with electric charges and string dipoles, in
theories of gravity coupled to a 2-form or 3-form field strength and to
a dilaton with arbitrary coupling. The method allows to describe not
only black holes with large angular momenta, but also other regimes that
include charged black holes near extremality with slow rotation. We
construct explicit examples of electric rotating black holes of
dilatonic and non-dilatonic Einstein-Maxwell theory, with horizons of
spherical and non-spherical topology. We also find new families of
solutions with string dipoles, including a new class of prolate black
rings. Whenever there are exact solutions that we can compare to, their
properties in the appropriate regime are reproduced precisely by our
solutions. The analysis of blackfolds with string charges requires the
formulation of the dynamics of anisotropic fluids with conserved
string-number currents, which is new, and is carried out in detail for
perfect fluids. Finally, our results indicate new instabilities of
near-extremal, slowly rotating charged black holes, and motivate
conjectures about topological constraints on dipole hair.

\noindent

%\keywords{AdS/CFT correspondence, black holes, fluid dynamics}

\vfill \setcounter{page}{0} \setcounter{footnote}{0}
\newpage

\tableofcontents

%\newpage

%%%%%%%%%%%%%%%%%%%%%%%%%%%%%%%%%%%%%%%%%%%%%%%%%%%%%%%%%%%%%%%%%%%%%%%%%%%%%%
%%%%%%%%%%%%%%%%%%%%%%%%%%%%%%%%%%%%%%%%%%%%%%%%%%%%%%%%%%%%%%%%%%%%%%%%%%%%%%
\section{Introduction}

Charged black holes in dimensions higher than four play an important
role in supergravity and in string theory, and large families of them
are explicitly known. If they are also rotating, we expect their
dynamics to become at least as rich as for neutral black holes
\cite{Emparan:2008eg,Myers:1986un}. However, the number of known exact
solutions with both charge and rotation is considerably smaller than in
the absence of either of them. Restricting to theories with a single
gauge field, possibly coupled to a dilaton (and without a cosmological
constant), the known exact black hole solutions\footnote{We only
consider asymptotically flat black holes that are regular on and outside
a connected event horizon.} with electric charge reduce, in five
dimensions, to the BMPV solution and its non-extremal extension
\cite{bmpv} plus the black rings of
\cite{Elvang:2004rt,Elvang:2003yy,Elvang:2003mj,Elvang:2004xi} (with a
Chern-Simons term). In $D>5$ there are the charged black holes that can
be obtained by U-duality-type transformations of the Myers-Perry black
holes \cite{Llatas:1996gh}, but these only allow a few values of the
dilaton coupling, in particular the Kaluza-Klein value
\cite{Kunz:2006jd}, but never zero coupling. In fact, a solution whose
exact form is conspicuously absent is the natural higher-dimensional
extension of the Kerr-Newman solution, namely a rotating charged black
hole of the Einstein-Maxwell theory, without dilaton coupling
nor Chern-Simons terms.

Given the plethora of novel black hole geometries and topologies
uncovered in recent years, many new classes of charged rotating
solutions can be anticipated. Some progress has been achieved
perturbatively, by adding a small rotation to a charged static black
hole \cite{perturbhs}, or a small charge to a rotating Myers-Perry black
hole or black ring \cite{perturcharge}. Numerical analysis has also been used to find
five-dimensional black holes with one angular momentum, or
odd-dimensional black holes with equal-magnitude angular momenta
\cite{numer}. Note that none of these black holes reach an ultraspinning
regime, and their horizons always have spherical topology.

In this paper we follow another approach that also produces approximate
solutions, but whose starting point is completely different. Neither the
amount of charge nor the rotation of the black holes need be small.
Indeed these solutions can be in the ultraspinning regime characteristic
of higher-dimensional black holes, but also in other regimes where the
mass, charge, and angular momenta enter in different proportions. A
regime that is particularly new for the method we use is one where the
charge is close to extremality and the rotation is very slow. Both
spherical and non-spherical horizon topologies appear naturally in all
regimes. And perhaps most remarkably, to leading order in the
approximation the solutions are obtained very easily in analytic form in
many cases of interest, and the calculation of their physical properties
is straightforward and often very simple.

The method also lends itself to the study of black holes with dipoles.
In $D\geq 5$ a black hole can act as an electric source for a
$(q+2)$-form field strength $H_{[q+2]}$, in a theory of the form
\beq\label{theory} 
I=\frac{1}{16\pi G} \int d^Dx \sqrt{-g}\left(
R-2(\nabla\phi)^2-\frac{1}{2(q+2)!} e^{-2a\phi}H_{[q+2]}^2 \right)\,.
\eeq 
However, if the spacetime is asymptotically flat, only when $q=0$ can
the black hole possess a conserved monopolar electric charge. When
$q\geq 1$ no conserved charge can be associated to its field. Instead,
the black hole carries a dipole. The first such solutions were found in
\cite{Emparan:2004wy} in the form of five-dimensional dipole black
rings. In fact they remain the only known dipole black holes of
\eqref{theory} with $q\geq 1$,\footnote{This refers to dipole black
holes with a single, connected horizon, and no charge. Refs.~\cite{yaz}
present multi-black holes of this theory with dipoles. The five-dimensional
rotating black holes and rings  of
\cite{Elvang:2003yy,Elvang:2003mj,Elvang:2004xi,Llatas:1996gh,Kunz:2006jd} 
are solutions of of \eqref{theory} with magnetic charge and electric
dipole when $a\neq 0$.} but it is to be expected that large classes of
dipole black holes exist, in particular in $D\geq 6$.\footnote{However,
there are strong restrictions on the possible existence of
\textit{static} black holes with dipoles \cite{Emparan:2010ni}.} The
only approach developed so far that allows to investigate them
systematically is the one employed here (other particular instances of
the method have been studied in \cite{Grignani:2010xm}, and will also be
presented in \cite{chargedbfolds}).

In this paper we only consider the theories \eqref{theory} with $q=0$ or
$q=1$, so the black holes have 0-brane (\ie particle) charge, or
1-brane (\ie string) dipole\footnote{Our solutions are dual to magnetic
sources of fields $H_{[D-q-2]}$, but other than this we do not consider
magnetic charges, nor dyonic solutions.}. A main reason is that this
allows us to perform a very complete and explicit study of generic
properties of the solutions. The dilaton coupling $a$
remains arbitrary, so in particular our study includes the pure
Einstein-Maxwell theory $q=0$, $a=0$. Values $q>1$ will be investigated
elsewhere.

The method we employ is the \textit{blackfold approach} of
\cite{Emparan:2009cs,Emparan:2009at}. The black hole is constructed as a
black $p$-brane whose worldvolume wraps a submanifold in a background
spacetime. The black brane is thin, in the sense that the horizon size
$r_0$ in directions transverse to the worldvolume is much smaller than
any characteristic length scale $R$ along the worldvolume. To leading
order in $r_0/R$, the brane is treated as a `probe' whose backreaction
on the background is neglected. At higher orders, backreaction effects
can be systematically included by performing a matched asymptotic
expansion \cite{Emparan:2007wm}. However, the leading order
approximation often gives enough non-trivial information about
stationary solutions. Using this approach
refs.~\cite{Emparan:2009cs,Emparan:2009vd} have presented large classes
of new neutral black holes.

Charges can be naturally incorporated in this framework. The starting
point now are black $p$-brane solutions of the theories \eqref{theory}
with charges of $q$-branes. We can think of the $p$-brane as having a
charge density of objects that extend along $q$ spatial directions,
`dissolved' in its worldvolume. Therefore, when $q=0$ the $p$-brane has
pointlike electric charges on its worldvolume, and if it wraps a compact
submanifold of the background, we obtain a black hole with electric
charge. When $q>0$, a blackfold with compact worldvolume gives a black
hole with an electric $q$-brane dipole, as discussed above. Since the
charge introduces one more scale in the problem, the regime of
applicability of the method has to be reassessed. We find that when the
0-brane charge is sufficiently large, the black hole need not be
ultraspinning in order to resemble locally a black brane. Due to the
cancellation of forces in the charged extremal limit, one can construct
brane-like distributions of `charged dust'. These extremal configurations
are singular, but they can be thermally excited above extremality to
exhibit regular horizons, and the (small) tensions that appear are then
balanced with suitable slow rotations.

The equations that describe the dynamics of blackfolds decompose into
two (coupled) sets: intrinsic equations, which take the form of
fluid-dynamics on the worldvolume, and extrinsic equations, which
characterize the embedding of the worldvolume in the background. The
inclusion of charges has the effect that the intrinsic worldvolume
theory is that of a $p+1$-dimensional fluid with a conserved $q$-brane
number current, with $q\leq p$. The case $q=0$ is a familiar one --- a
fluid with a conserved particle number --- but fluids with string or
brane currents do not seem to have been studied before, at least not
relativistic ones\footnote{Liquid crystals are examples of fluids with
string-like excitations, but we have not found any useful way to
capitalize on their vast literature.}. In this paper we develop in
detail the general theory of perfect fluids with $q$-brane currents for
$q=0,1$. Blackfolds with $q=p$ (plus other examples) will be developed
in \cite{chargedbfolds} and the study of generic $q$ will be presented
elsewhere.

We perform an analysis of stationary solutions that completely
solves the intrinsic equations in a very general and explicit manner,
and which paves the way for the construction of new black holes with
charges and dipoles. The extrinsic equations are in general differential
equations, but, in analogy to \cite{Emparan:2009vd}, we find several
classes of simple worldvolume geometries that can be solved
algebraically. These are:

\paragraph{Charged black holes:}
\begin{itemize}

\item Charged black holes with horizon topology $S^{D-2}$ in $D\geq 6$,
with rotation on $s$ independent planes (with $1\leq s< \frac{D-3}{2}$) and
arbitrary dilaton coupling. These are the generalization of the
Myers-Perry solutions to include electric charge, in the regime where
their horizons are flattened along the planes of rotation (which can,
but need not, be ultraspinning). In particular, they include the elusive
electric rotating black holes of the Einstein-Maxwell theory ($a=0$),
which for the first time are described in these regimes. They also
include the previously known charged rotating black holes in
Kaluza-Klein theory \cite{Llatas:1996gh,Kunz:2006jd}, whose properties
we show are accurately reproduced in the appropriate regime. We shall
argue that in the blackfold regime all these black holes are unstable even
when they are close to extremality and slowly rotating.

\item Charged black holes with horizon topology $\Pi_{p_i\in
\mathrm{odd}}S^{p_i}\times S^{D-\sum_i p_i-2}$. These are the natural
charged generalization of the solutions in
\cite{Emparan:2009cs,Emparan:2009vd}. Besides the exact five-dimensional
black rings of \cite{Elvang:2003yy,Elvang:2003mj,Elvang:2004xi}, which
we reproduce correctly when the rings are thin, no
other construction of them
exists.

\end{itemize}

\paragraph{String-dipole black holes:}
\begin{itemize}
\item Black holes with string dipole in $D\geq 5$, with horizon topology
$\Pi_{p_i\in \mathrm{odd}}S^{p_i}\times S^{D-\sum_i p_i-2}$. The
properties of the exact five-dimensional dipole rings of
\cite{Emparan:2004wy} are accurately reproduced in the ultraspinning
limit.
\item A qualitatively different class of black rings with string dipole
in $D\geq 6$. They are constructed as blackfolds with annulus-shaped
($S^1\times I$) and more generally solid-ring-shaped ($S^1\times B_{2k-1}$)
worldvolumes. Their horizon topology is ring-like, $S^1\times S^{D-3}$,
but their geometry is unlike the previous solutions: the $S^{D-3}$ is
pancaked along the planes on which the strings lie, so the shape
resembles that of a `prolate ring'\footnote{We call a shape {\em
prolate} if the directions in which it is lengthened include directions
orthogonal to the rotation planes (which are typically lengthened too).
This effect, induced by the repulsion among parallel strings, is to be contrasted to
the flattening due to centrifugal forces.}. There are also blackfolds
with worldvolumes $I\times S^{2k-1}$
which give rise to similarly prolate versions of dipole black holes with
product-of-spheres horizons.
\end{itemize}

This is far from being an exhaustive classification, instead it is
mainly intended to illustrate the power of the method, in particular, of
the general solution of the intrinsic equations, and the new regimes
near extremality that we can also investigate.

Finally, it must be noted that although we have written the theories
\eqref{theory} without Chern-Simons terms, which are present in
particular in five-dimensional minimal supergravity, in many cases the
solutions we construct also apply with them. When the rotation occurs in
a direction parallel to the brane current, as \eg in dipole black
strings and rings, it generates no magnetic field and the Chern-Simons
term is irrelevant (dipole black rings in five dimensions are often
regarded as magnetic solutions of the $q=0$ theory \eqref{theory}, and
in this case it is the absence of induced electric sources that makes
the Chern-Simons term unimportant to them). 

When the rotation occurs in
directions transverse to the current, as is necessarily the case for
black holes with electric 0-brane charges, and possibly when $q<p$, then
the rotation induces a magnetic dipole moment and then it makes a
difference whether the Chern-Simons term is present or not. Even if at
first sight this effect might seem subleading in the blackfold
expansion, it is known that in the case of five-dimensional black rings
it leads to rotation in the $S^2$ and to a regularity condition of global
type, namely absence of Dirac-Misner strings, which requires keeping
both the electric charge and the magnetic dipole even at arbitrarily
large, but finite, ring radius. This is feasible in the blackfold
framework, but is outside the scope of this article.

\paragraph{Outline of the paper.} Section~\ref{sec:genform} develops the
general formalism for blackfolds with $q$-brane charges. Although the
main focus is on $q=0,1$, we obtain the effective stress-energy tensor
for black $p$-branes with generic $q$-brane charges. In
section~\ref{sec:statbfs} we solve in a complete and explicit manner the
intrinsic, fluid-dynamical equations for stationary configurations with
$q=0,1$. We present a simple action principle for the extrinsic
dynamics, give expressions for the physical properties of the solutions,
and obtain their thermodynamics. In section~\ref{sec:range} we discuss
the applicability of the method and identify the new regime of
near-extremal, slowly rotating charged blackfolds. The subsequent
sections are devoted to the construction and study of specific solutions
obtained by applying the general study of sec.~\ref{sec:statbfs}.
Sec.~\ref{sec:0charge} focuses on blackfolds with 0-brane charge, first
with disk and even-ball worldvolumes, then with odd-sphere worldvolumes.
Sec.~\ref{sec:1charge} analyzes blackfolds with string dipole, and in
particular we find that disk solutions are not possible, instead a new
class of rings appears. We conclude in section~\ref{sec:outout} with a
discussion of our results.

We collect a number of technical details in a series of appendices.
Appendix~\ref{app:branesol} constructs the black $p$-brane solutions of
\eqref{theory} with $q$-brane charge as uplifts of Gibbons-Maeda black
holes, and appendix~\ref{app:charges} computes their stress-energy
tensor and all physical magnitudes of interest, and discusses their
thermodynamics. Appendix~\ref{app:comparison} compares the ultraspinning
regime of the exact solutions for rotating black holes with Kaluza-Klein
electric charge \cite{Kunz:2006jd}, and for black strings and rings with
string charge \cite{Emparan:2004wy} with the solutions in
secs.~\ref{sec:0charge} and \ref{sec:1charge}. In all cases the
agreement is perfect.

%\smallskip

\paragraph{Notation.} We follow ref.~\cite{Emparan:2009at}:

\textit{Spacetime (background)} objects: coordinates $x^\mu$, metric
$g_{\mu\nu}$, connection $\nabla_\mu$, with
$\mu,\nu,\rho,\ldots=0,\dots, D-1$; $h_{\mu\nu}$ is the first
fundamental form of the embedding of the worldvolume $\mc W_{p+1}$ in
the background; $\perp^\mu{}_\nu$ is the projector orthogonal to $\mc W_{p+1}$.

\textit{Worldvolume} objects: coordinates $\sigma^a$, with
$a,b,c,\ldots=0,\dots,p\,$; the metric
$\gamma_{ab}$ induced on the worldvolume is the pullback
of $h_{\mu\nu}$; its compatible connection, $D_a$, is
the pullback of $h_\mu{}^\nu\nabla_\nu$.

\textit{Worldvolume current} objects are denoted with hats: $\hat h_{ab}$ is the first
fundamental form of the current worldline/sheet $\mc C_{q+1}$ inside $\mc W_{p+1}$;
$\hat\perp_{ab}=\gamma_{ab}-\hat h_{ab}$ the projector orthogonal to
$\mc C_{q+1}$; $|\hat h|^{1/2}$ is the area element of $\mc C_{q+1}$.

%$\mc W_{p+1}$ is the blackfold worldvolume; $\mc B_p$ its spatial
%section; $\mc C_2$ are the worldsheets of strings dissolved in the
%worldvolume.
For a $p$-brane in $D$ spacetime dimensions we define
\eq
n=D-p-3\geq 1\,.
\eeq
Instead of the dilaton coupling $a$ it is more convenient to use $N$
such that
\eq\label{Ndef}
a^2=\frac{4}{N}-\frac{2(q+1)(D-3-q)}{D-2}\,.
\eeq
%\eq\label{xidef}
%N=\frac{4(D-2)}{2(q+1)(D-3-q)+(D-2)a^2}\,.
%\eeq
$N$ is preserved under dimensional reduction and is a
real number in general, but in many instances of interest in
supergravity and string theory it is an integer.

When we talk about black holes, as opposed to
infinitely extended black branes, we find more appropriate to use the
term `charged black hole' for the solutions with 0-brane charge, and
`dipole black holes' for the ones with $q\geq 1$ brane dipole.

%%%%%%%%%%%%%%%%%%%%%%%%%%%%%%%%%%%%%%%%%%%%%%%%%%%%%%%%%%%%%%%%%%%%%%%%%%%%%%
%%%%%%%%%%%%%%%%%%%%%%%%%%%%%%%%%%%%%%%%%%%%%%%%%%%%%%%%%%%%%%%%%%%%%%%%%%%%%%
\section{Blackfolds with electric charge and with string dipoles}
\label{sec:genform}

%%%%%%%%%%%%%%%%%%%%%%%%%%%%%%%%%%%%%%%%%%%%%%%%%%%%%%%%%%%%%%%%%%%%%%%%%%%%%%

Refs.~\cite{Emparan:2009cs,Emparan:2009at} show that whenever there are
two separate scales along the
horizon of a black hole, one can effectively
describe its physics by integrating out the smaller scale, leading to a
thin black $p$-brane curved on a submanifold ${\mathcal W}_{p+1}$
embedded in a background spacetime. Its dynamics is determined by a
stress tensor $T_\munu$ that is supported on its
worldvolume. The geometry of the submanifold ${\mathcal
W}_{p+1}$ is completely defined by the combination of its induced
worldvolume metric
\beq
\ga_{ab}=\p_aX^\mu\p_bX^\nu g_{\munu}
\eeq
that fixes
the intrinsic geometry, and of the extrinsic curvature tensor
$K_\munu{}^\rho$ that encodes the shape of the embedding. The projector
to the tangent space of ${\mathcal W}_{p+1}$ is given by the first
fundamental form
\beq
h^\munu=\p_aX^\mu\p_bX^\nu\ga^{ab},
\eeq
while the
projector to orthogonal directions is
\beq
\perp_\munu=g_\munu-h_\munu\,.
\eeq
Then $K_\munu{}^\rho=h_\mu{}^\la h_\nu{}^\si\nabla_\si h_\la{}^\rho$,
and its trace is the mean curvature vector $K^\rho=h^\munu
K_\munu{}^\rho$.

The collective variables describing a neutral blackfold
are the embedding functions $X^\mu(\si)$ characterizing the manifold
${\mathcal W}_{p+1}$, the thickness $r_0(\si)$ of the horizon, and the
local boost field $u^a(\si)$ satisfying $\ga_{ab}u^au^b=-1$. All
collective variables depend on the worldvolume coordinates $\si^a$,
$a=0,\ldots,p$, and are allowed to vary over a length scale $R$
much longer than the scale $r_0$, set by the brane thickness, of the
physics we have integrated out (we will be more precise about this point
in sec.~\ref{sec:range}).
The result is a long-wavelength effective theory of the black hole
defined by the effective stress tensor $T_\munu$ of the brane.

When the black hole is charged, there are additional collective
variables that combine into an effective charge current $J$ that flows
on the blackfold worldvolume $\mc W_{p+1}$. In general this will be a
$(q+1)$-form describing a conserved $q$-brane current in $\mc
W_{p+1}$.\footnote{Note that there is no gauge field on the worldvolume
theory, and the worldvolume current $J$ is a global one. This is just
like in the AdS/CFT correspondence, with which the blackfold approach
has many similarities.} The collective variables that characterize it
are a $q$-brane charge density $\mc Q(\si)$ and a set of worldvolume
vectors for the orientation (polarization) of the $q$-branes inside the
worldvolume. The applicability of the blackfold approach in the
presence of charges will be discussed later.

\subsection{Fluid dynamics with particle- and string-number currents}

In this article we restrict ourselves to blackfolds which carry
$0$-brane (particle) and $1$-brane (string) charges diluted in their
worldvolume. The formalism for perfect fluids with a conserved
$q$-brane current with $q=0,1$, which we develop next, remains simple
enough.

When $q=0$ this is the familiar fluid dynamics with a conserved particle
number (like \eg baryon number). The
stress-energy tensor of the fluid, $T_{ab}$, has a unique unit-normalized timelike
eigenvector $u$ whose eigenvalue $-\vep$ defines the energy density. Spatial
isotropy then implies that
\beq\label{stress0}
T_{ab}=\vep u_a u_b +P (\gamma_{ab}+u_a u_b)
\eeq
with pressure $P$. As is well known, in the absence of dissipative effects
(specifically, no particle diffusion) the particle current must be
proportional to $u$,
\beq\label{J1}
J_a=\mathcal Q\,u_a\,,
\eeq
where $\mathcal
Q$ is the charge density on the fluid.

When $q=1$, we can contract the timelike eigenvector $u$ of
$T_{ab}$ with the string two-form current
$J_{ab}$ to find the spacelike vector $u_a
J^{ab}$, which is orthogonal to $u$. Normalizing this vector $v$ to one, so that
\beq\label{uv}
-u^2=v^2=1\,,\qquad u\cdot v=0\,,
\eeq
we can then write
\beq\label{J2}
J_{ab}=\mc Q (u_a v_b-v_a u_b)\,,
\eeq
which defines the string charge density $\mc Q$.

The vector $v$ breaks the spatial isotropy of the fluid and characterizes
the directions along which the (dissolved) strings lie. Again, if dissipative
currents are absent, $v$ must be an eigenvector of $T_{ab}$.
The fluid is isotropic in the spatial
directions transverse to it, so
\beq\label{Tabanis0}
T_{ab}=\vep u_a u_b +P_\| v_a v_b +P_\perp\lp \gamma_{ab}+u_a u_b- v_a v_b\rp\,.
\eeq
Note that this analysis easily allows the simultaneous presence of 0- and
1-brane-number currents, but in this paper we will not study such systems.

We can unify the description of fluids with either of
these currents if we
introduce the projector onto the space parallel to the particle/string
worldline/sheet,
\eq
\hat h_{ab}=-u_a u_b+q\,v_a v_b\qquad (q=0,1)\,,
\label{hhat}\eeq
and onto directions in $\mc W_{p+1}$ orthogonal to it,
\eq\label{hath}
\hat\perp_{ab}=\gamma_{ab}-\hat h_{ab}\,.
\eeq
Then
\beq\label{Tabanis}
T_{ab}=(\vep+P_\|) u_a u_b  +(P_\|-P_\perp) \hat h_{ab}
+P_\perp \gamma_{ab}\,.
\eeq
If we also introduce the volume form $\hat V$ on the worldline/sheet,
\beq
\hat V=\left\{ \begin{array}{ll}
u\vphantom{\wedge v}& \quad\mathrm{for}~q=0\,\\
u\wedge v&\quad\mathrm{for}~q=1\,,
\end{array} \right.
\eeq
then
\beq
J=\mc Q\; \hat V\,.
\eeq

The difference between the pressure in directions orthogonal to the
strings and parallel to them is due to the energy density of the strings
(essentially their tension), given by $\Phi\mc Q$, with $\Phi$ the
string chemical potential, so in general
\beq\label{pressdiff}
P_\perp-P_\|=\Phi\mc Q\,.
\eeq
In addition to this, thermodynamic equilibrium is satisfied locally on
the worldvolume so we have the first law
\beq\label{firstlaw}
d\vep=\mc T ds+\Phi d\mc Q\,,
\eeq
where $\mc T$ is the local temperature, and also the thermodynamic Gibbs-Duhem
relations
\beq\label{GDint}
\vep+P_\perp=\mc T s+\Phi\mc Q\,,
\eeq
\beq\label{GDdiff}
dP_\perp=sd\mc T+\mc Q d\Phi\,,\qquad
dP_\|=sd\mc T-\Phi d\mc Q \,.
\eeq

Counting the independent collective variables, we find $p$ independent
components for $u$, $p-1$ components for $v$ when $q=1$, plus the energy density,
charge density, and the
pressures, which add up to $(q+1)p+3$ variables. They
must satisfy
the current continuity equations
\beq\label{curcont}
d\ast J=0
\eeq
($\ast$ is the Hodge dual on $\mc W_{p+1}$)
and the fluid equations
\beq\label{emcons}
D_aT^{ab}=0\,.
\eeq
This is a set of $(q+1)p+2$ equations, which, when supplemented with the
equation of state that determines $\vep$ for given pressures and charge
density, are enough
to determine the system.

We proceed now to write down these equations more
explicitly. We denote
\beq
\dot u\equiv u^a D_a u\,,\qquad D_v v\equiv v^a D_a v\,.
\eeq

\paragraph{Intrinsic equations with particle and string currents.}

The exterior product of the current continuity equation
\eqref{curcont} with
$\ast\hat V$ gives
\beq
\ast \hat V \wedge d\ast \hat V=0\,.
\eeq
This equation expresses,
through Frobenius' theorem, the existence of integral
$(q+1)$-dimensional submanifolds parallel to $\hat h_{ab}$.
This
is trivial for particle currents, but for
string currents it implies the integrability of the submanifolds
${\mathcal C_2}\subset\mathcal
W_{p+1}$ whose tangent space is spanned by $(u,v)$. These submanifolds
are the worldsheets of the strings carrying the charge $\mc Q$. In
index notation the equation is
\eq
\hat\perp_{ab}\lb u,v\rb^b=0\,.
\label{comm}\eeq

The remaining current continuity equations can be written as
\beq
D_a(\mc Qu^a)=0
\eeq
for $q=0$ and
\beqa
D_a(\mc Q u^a)+\mc Q u^a D_v v_a&=&0\,,\label{currv}\\
D_a(\mc Q v^a)-\mc Q v^a \dot u_a&=&0\label{curru}\,,
\eeqa
for $q=1$.
Using these and the local thermodynamics equations
%\eqref{pressdiff}, \eqref{firstlaw}, \eqref{GDint}, \eqref{GDdiff},
we can write the fluid
equations \eqref{emcons} as the conservation of entropy
\beq\label{entcons}
D_a(su^a)=0\,,
\eeq
and the (Euler) force equations
\beq\label{perpdT}
\hat\perp^{ab}s\mc T\lp
\dot u_b+\p_b\ln\mc T\rp
-\mc Q\Phi\lp
\hat K^a-\hat\perp^{ab}\p_b\ln\Phi\rp=0\,
\eeq
and
\beq\label{vdT}
\lp\hat h^{ab}+u^au^b\rp\lp\dot u_b +\p_b\ln\mc T\rp=0%\,.
\eeq
(which is trivial when $q=0$). Here we have introduced
the mean curvature of the worldlines/sheets embedded in
$\mathcal W_{p+1}$,
\eq
\hat K^a=\hat h^{bc}D_b \hat h_c{}^a=
-\hat\perp^a{}_b\lp \dot u^b-q\,D_v v^b\rp\qquad (q=0,1)\,.
\label{khat}\eeq

\subsection{Extrinsic dynamics}

The theory of blackfolds has a second set of equations, extrinsic
ones, that describe the
dynamics of the embedding of the $p$-brane worldvolume in the
$D$-dimensional background spacetime. The $D-p-1$ independent transverse
coordinates $X(\si)$ that characterize this embedding are determined by
solving Carter's equations \cite{Carter:2000wv},
\eq
T^\munu K_\munu{}^\rho=
\frac1{(q+1)!}\perp^\rho{}_\sigma J_{\mu_0\dots\mu_q} H^{\mu_0\dots\mu_q\sigma}\,,
\label{carter0}\eeq
where $H_{[q+2]}$ is the background field strength that couples electrically to
the $q$-brane charge.
In this article we
shall be concerned only with situations where such fields are absent, so
\eq
T^\munu K_\munu{}^\rho=0\,.
\label{carter}\eeq

The stress tensor is the push-forward
$T^{\mu\nu}=\partial_a X^\mu
\partial_b X^\nu T^{ab}$ of the $q$-brane-charged fluid stress tensor
\eqref{Tabanis}. Then
\eqref{carter} becomes (\eg see eq.~(2.2) of \cite{Emparan:2009vd})
\beq\label{ext1}
P_\perp K^\rho=-\perp^\rho{}_\mu\left(
s\mc T\dot u^\mu-\Phi \mc Q \hat K^\mu\right)\,,
\eeq
where
\eq
\hat K^\mu=\hat h^{\nu\si}\nabla_\nu \hat h_\si{}^\mu
=\hat h^{\nu\si}K_{\nu\si}{}^\mu\,
\label{khatmu}\eeq
is the mean curvature of the worldlines/sheets of the particles/strings
embedded in the $D$-dimensional background spacetime, with the first
fundamental form $\hat h_{\mu\nu}$ obtained by pushing forward
\eqref{hhat}. Note that $\hat K^\mu$ has components both along
transverse and along parallel directions to $\mathcal W_{p+1}$, the
pull-back of the latter being \eqref{khat}.

\subsection{Effective stress-energy tensor for charged blackfolds}
\label{subsec:bfstress}

The next step is to compute the effective stress tensor for a black
$p$-brane with $q$-brane charge in its worldvolume. The expressions we
find in this section are valid for $0\leq q\leq p$, although in this
paper we only put them to use when $q=0,1$.

The effective stress tensor of the blackfold is defined as the quasi-local
stress tensor of Brown and York \cite{Brown:1992br} measured at a large distance
in directions transverse to the brane. We will restrict ourselves to the
leading order in a derivative expansion of $T^{ab}$, \ie a perfect
fluid. Dissipative corrections can be computed in the
blackfold formalism \cite{Camps:2010br}, but they do not
alter the properties of the stationary blackfolds that are the main
interest of this article.

To this leading order we must compute the
quasi-local stress tensor for the corresponding planar, equilibrium brane. A large
class of exact solutions of the theories (\ref{theory}),
describing black $p$-branes with diluted $q$-brane charges in
$D$-dimensional general relativity coupled to a $(q+1)$-form gauge
potential $B_{[q+1]}$ and a dilaton $\phi$ with dilaton coupling $a$, is
obtained in appendix~\ref{app:branesol}, see
eqs.~(\ref{branemetric})--(\ref{dilaton}). They
depend on two parameters, the radius $r_0$ and a charge
parameter $\alpha$. The quasi-local stress tensor they source is
computed in appendix~\ref{app:BY}. It is an anisotropic perfect fluid,
since the diluted $q$-brane
charges sort out privileged directions along which their worldvolumes
extend. Indeed, the net effect of these additional branes is to increase
the tension along the directions parallel to them. The energy
density $\varepsilon$, the pressure $P_\parallel$ in the $q$-brane
directions and the pressure $P_\perp$ in the transverse directions are
found to be
\eqa\label{bffluid1}
&&\varepsilon=\frac{\Omega_{(n+1)}}{16\pi G}r_0^n\lp n+1+n N\sinh^2\al\rp,\\
&&P_\parallel=-\frac{\Omega_{(n+1)}}{16\pi G}r_0^n\lp1+n N\sinh^2\al\rp,\qquad
P_\perp=-\frac{\Omega_{(n+1)}}{16\pi G}r_0^n.\label{bffluid2}
\eeqa
%leading to the equation of state
%\eq
%\varepsilon=-P_\parallel-nP_\perp\,.
%\eeq
Here, $\al$ is the charge parameter of the solution, linked to the
electric potential
%at the horizon, as measured with respect to infinity by
\eq
\Phi%=B_{ty^1\cdots y^{q}}(\infty)-B_{ty^1\cdots y^{q}}(r_0)
=\sqrt N\tanh\al\,,
\label{phi}\eeq
while the charge density, computed in appendix~\ref{app:charge}, is
\eq
\mathcal Q=\frac{\Omega_{(n+1)}}{16\pi G}n\sqrt{N}r_0^n\sinh\al\cosh\al\,.
\label{Qbrane}\eeq
The parameter $N$ was defined in \eqref{Ndef} (see also \eqref{Ndef2}).
Observe that all these densities depend on $n$ and $N$, but not on $p$
nor $q$. The reason is that under compactification of the $p$- and
$q$-brane directions these densities become the conserved charges of a dilatonic
black hole, which depend only on the number $n+3$ of dimensions that the
black hole lives in, and on the dilaton coupling through the parameter
$N$ that is invariant under dimensional reduction.

From these expressions we find the equation of state
\beq\label{eos}
\vep=-P_\|-n P_\perp\,,
\eeq
or using \eqref{pressdiff}, which is also satisfied, we can write it as
\beq
\vep-\Phi \mc Q =-(n+1)P_\perp\,.
\eeq

Finally, the local temperature and entropy density of this black brane are
\eq
{\mathcal T}=\frac n{4\pi r_0\lp\cosh\al\rp^{N}} ,\qquad
s=\frac{\Omega_{(n+1)}}{4G}r_0^{n+1}\lp\cosh\al\rp^{N}.
\label{Ts}\eeq

In the blackfold framework one promotes the constants $r_0$ and $\al$
defining the black brane solution to slowly varying functions $r_0(\si)$
and $\al(\si)$ of the worldvolume coordinates. Furthermore, a boost
along the worldvolume is introduced, which is taken to be characterized
by a slowly varying velocity $u^a(\si)$. For string charge, the
polarization vector becomes another collective variable, $v^a(\si)$.
A solution for the fluid is fully specified when we determine $u(\si)$ and
$v(\si)$, and a pair of worldvolume functions such as $(\mc
T(\si),\Phi(\si))$ or $(r_0(\si),\alpha(\si))$.

For the specific blackfold fluid given
by \eqref{bffluid1},
\eqref{bffluid2}, the stress tensor \eqref{Tabanis} is
\eq\label{Tabbf}
T_{ab}=\frac{\Om_{(n+1)}}{16\pi G}r_0^n\lp n\,u_au_b-\ga_{ab}
-nN\sinh^2\al\;\hat h_{ab}\rp
\eeq
and the extrinsic equation becomes
\beqa\label{exteqn1}
K^\rho=n\perp^\rho{}_\mu\lp \dot u^\mu -N\sinh^2\alpha\; \hat K^\mu\rp\,.
\eeqa

In particular, for
0-brane charge this is
\beqa\label{exteqn0}
K^\rho=n(1 +N\sinh^2\alpha)\perp^\rho{}_\mu\, \dot u^\mu\,.
\eeqa
We see that the effect of 0-brane charge dissolved on
the worldvolume is to \textit{decrease} the value of the
acceleration needed to
sustain a worldvolume of given mean curvature. This is because, for a
given energy density, the
0-brane charge decreases the effective tension on the worldvolume of the
black brane, thus making it easier to bend it. This has
significant consequences that we discuss in sec.~\ref{sec:range}.

For blackfolds with $q$-brane charge with $q>0$, the worldvolumes $\mc
C_{q+1}$ with positive mean curvature have the effect of
\textit{increasing} the acceleration needed to sustain a given curvature
of $\mc W_{p+1}$. This is in agreement with our previous observation
that, for a given transverse tension, $q$-brane charge increases the
tension along the directions in which the $q$-branes extend.

\subsection{Blackfolds with boundaries}
\label{subsec:bfbdries}

The equations that a charged blackfold must satisfy if its worldvolume
has boundaries are a simple extension of those found in
\cite{Emparan:2009at} for neutral blackfolds.
If $f(\si)$ is a function that defines the boundaries by
$f|_{\partial\mathcal{W}_{p+1}}=0$, then we must have
\beq
\left.u^\mu \partial_\mu f\right|_{\partial\mathcal{W}_{p+1}}=0
\eeq
at all boundaries.

Additionally, the string current conservation equation requires
\beq
\mc Q\, u^{[\mu}v^{\nu]}\partial_\mu\left. f\right|_{\partial\mathcal{W}_{p+1}}=0
\eeq
so
\beq\label{stringbdry}
\left.\mc Q\, v^{\mu}\partial_\mu f\right|_{\partial\mathcal{W}_{p+1}}=0\,.
\eeq
\ie the string current must remain parallel to the boundaries.

Conservation of the
stress-energy requires that in the absence of any
surface tension, the fluid pressure must
vanish on the boundary,
\beq
\left.P_\perp\right|_{\partial\mathcal{W}_{p+1}}=0\,.
\eeq
For a blackfold fluid this means that at the edge,
\beq\label{r0vanish}
\left. r_0\right|_{\partial\mathcal{W}_{p+1}}=0\,.
\eeq
It must be noted that this is a necessary condition for regularity of
the blackfold horizon, but it is not sufficiently understood yet under
what circumstances it may also be sufficient.

The meaning of \eqref{r0vanish} for charged blackfolds will be further
clarified after we analyze stationary solutions.

%%%%%%%%%%%%%%%%%%%%%%%%%
\section{Stationary charged blackfolds}
\label{sec:statbfs}

The general study of charged stationary blackfolds can be performed following
closely the lines of \cite{Emparan:2009at}. The intrinsic equations can
be solved completely and then the extrinsic equations can be
conveniently encoded in a simple variational principle. Explicit
expressions for the physical magnitudes can be found which satisfy the
laws of black hole thermodynamics.

\subsection{Solving the intrinsic fluid equations}

Stationarity implies that
the black brane has a Killing horizon associated to a timelike Killing
vector $k$ with a surface gravity $\kappa$. The velocity field of
the fluid is set to be proportional to this Killing vector,\footnote{Do
not confuse the redshift factor $\gamma$ with the metric induced in the
worldvolume $\gamma_{ab}$.}
\eq
u^a=\ga k^a,\qquad \ga=1/\sqrt{-k^2}\,.
\label{uk}\eeq
This is a general result, proven in \cite{Caldarelli:2008mv} for general
stationary fluid configurations, and extended in \cite{Emparan:2009at}
to neutral blackfolds.
In the presence of charges, we assume again this form
of the velocity field, which implies that
\beq\label{nablau}
D_{(a}u_{b)}=u_{(a}\p_{b)}\ln\ga\,,
\eeq
and therefore the expansion and shear of $u$
vanish and its acceleration is
\eq
\dot u_a=-\p_a\ln\ga.
\label{acceleration}\eeq

Stationarity of the full configuration requires that any
physical quantity
characterized by a tensor $\mathbb{T}$ be constant
along the Killing flow, so its Lie derivative vanishes
\beq\label{statT}
\mathcal{L}_k \mathbb{T}=0\,.
\eeq
Note that \eqref{uk} implies $\mathcal{L}_k u=0$ and $\mathcal{L}_k
\gamma=0$.

In order to solve the intrinsic
equations for a charged fluid we use \eqref{acceleration} and that, in
general, the extrinsic curvature vector of the worldlines/sheets inside $\mc
W_{p+1}$ is
\eq
\hat K^a=
-\hat\perp^{ab}\p_b\ln\lp|\hat h|^{1/2}\rp\,,
\eeq
where $|\hat h|^{1/2}$ is the area element of the worldlines/sheets. Then
\eqref{perpdT} and \eqref{vdT} become
\beq\label{perpdT2}
s\mc T\hat\perp^{ab}\p_b\ln\frac{\mc T}{\ga}
+\mc Q\Phi\hat\perp^{ab}\p_b\ln\lp |\hat h|^{1/2}\Phi\rp=0
\eeq
and
\beq\label{vdT2}
\lp\hat h^{ab}+u^au^b\rp\p_b\ln\frac{\mc T}{\ga}=0\,.
\eeq
In principle these equations would allow a dependence of
$\mc T$ and $\Phi$ on the coordinate along $u$ (\ie time), but this would be
incompatible with stationarity.
We solve the equations by taking
\eq\label{constT}
{\mathcal T}(\si)=\ga T\,,
\eeq
where $T$ is an integration constant with the
interpretation of global temperature of the blackfold, and
\beq\label{Phisol}	
\Phi(\si)=\frac{\phi(\si)}{|\hat h|^{1/2}}\,,
\eeq
where $\phi(\si)$ can vary only along spatial directions parallel to
the current but orthogonal to $u$,
\beq
(\hat\perp^{ab}-u^a u^b)\p_b\phi(\si)=0\,.
\eeq
Then if we integrate over the spatial
directions along the current we find a constant 
\beq\label{totpot}
\Phi_H=\int_{\mc C_q}d^q\sigma \phi(\sigma)=\int_{\mc C_q}d^q\sigma |\hat h|^{1/2}\Phi(\si)
\eeq 
that may be regarded as the global potential.\footnote{The necessity
to integrate the potential along
 the worldvolume directions parallel to the current for obtaining the
condition for stationarity has been recognized previously in
\cite{chargedbfolds}.} 
%The factor $2\pi$ for $q=1$ is simply introduced
%for later convenience (our conventions for overall factors differ
%slightly from \cite{Emparan:2004wy}).

For 0-brane currents $\phi=\Phi_H$ fixes the local electric potential
$\Phi(\sigma)$ on the blackfold. Thus, given the Killing vector $k$ and
the constants $T$ and $\Phi_H$ all the collective variables for the
fluid are determined. This is therefore a complete and general solution
to the intrinsic equations for a stationary blackfold with 0-brane
charge.

For the fluid with string-number current the solution is not
complete yet, as we still need to find $v$ that solves
\eqref{comm}, and then fully specify $\Phi$.
In order to complete the solution we use the result \cite{Hollands:2006rj} that for a
stationary (but not static) black hole, there is a spacelike Killing
vector $\psi$ that commutes
with $k$,
\beq
[\psi,k]=0\,.
\eeq
Then we construct its component orthogonal to $k$
\beq
\zeta=\psi-\frac{\psi^a k_a}{k^2}k
\label{zeta}\eeq
and assume that $\zeta$ is spacelike over the blackfold worldvolume.
If $\zeta$ were found to become timelike on a region of the
worldvolume, then that region should be excluded and the blackfold would
have a boundary.

Note that $\psi^a k_a/k^2$ is a function that remains constant along the
two Killing directions
$\psi$ and $k$, but
which can
vary in directions transverse to them. So in general $\zeta$ is not a
Killing vector, but nevertheless it satisfies
\beq
\zeta^a D_{(a}\zeta_{b)}=0\,,\qquad k^a D_{(a}\zeta_{b)}=0\,,\qquad
D_a\zeta^a=0\,.
\eeq
If we take
\beq\label{vsol}
v^a=\frac{\zeta^a}{|\zeta|}\,,
\eeq
then
\beq
[u,v]=0
\eeq
so eq.~\eqref{comm} is satisfied.
Additionally,
\beq\label{Dva}
D_a v^a=0\,,\qquad D_v v_a =-\p_a\ln|\zeta|\,,\qquad v^a \dot u_a=0
\eeq
and eq.~\eqref{nablau} implies that $v^a D_v u_a=0$, or equivalently,
$u^a D_v v_a=0$. With these
\eqref{curru} becomes
\beq\label{vdQ}
v^a\p_a \mc Q=0\,.
\eeq
Together with \eqref{constT} this implies that $\Phi$ cannot vary along
$v$, so in \eqref{Phisol} we set $\phi$ to be a constant. If we assume
that the orbits of $\psi$ are closed we can normalize it so that the
periodicity is $2\pi$. The integration in \eqref{totpot} only involves
the cyclic direction along $\psi$, so
$\Phi_H=2\pi\phi$.

The area element on $\mc C_{q+1}$ is
\eq\label{hathsol}
|\hat h|^{1/2}=
\left\{ \begin{array}{ll}
\gamma^{-1}& \quad\mathrm{for}~q=0\,\\
|\zeta||k|=|\zeta|/\gamma&\quad\mathrm{for}~q=1\,,
\end{array} \right.
\eeq
and then
we can write the solution for the potential, both for $q=0,1$, as
\beq\label{refPhi}
\Phi(\si)=\frac1{(2\pi)^q}\frac{\Phi_H}{|\hat h(\si)|^{1/2}}\,.
\eeq

Inverting
(\ref{phi}) and (\ref{Ts}) we find
\eq\label{sol1brane}
r_0(\si)=\frac{n}{4\pi T}\ga^{-1}\lp1-
\frac1{(2\pi)^{2q}}\frac{\Phi_H^2}{N|\hat h|}\rp^{N/2},\qquad
\tanh\al(\si)=\frac1{(2\pi)^{q}}\frac{\Phi_H}{\sqrt{N}|\hat h|^{1/2}}\,.
\eeq
It is now clear that the equations for $s$ and $\mc Q$ are all solved.
Stationarity of all physical quantities is also guaranteed.

For $q=0$ this solution is general, while for $q=1$ the only 
stationary solutions that are not included in this analysis are
such that the spatial direction
of the current is not aligned with a vector constructed out of Killing vectors in the
form \eqref{zeta} (possibly because $\psi$ may not exist, in which case
the blackfold is necessarily static). In this case they must still
satisfy \eqref{constT}
and \eqref{Phisol}, but we do not have a complete explicit form for $v$
or $\Phi$. Such solutions may be of interest but we will not investigate
them in this paper.

Observe that our general stationary solution
\eqref{Phisol}
\eq
\hat K^a=\hat\perp^{ab}\p_b\ln\Phi\,,
\label{KC}\eeq
says that the gradient of
the chemical potential balances the stress due to
the mean curvature of the dissolved $q$-branes. When $q=0$ the latter
only involves the acceleration of the charged particles, but when $q=1$
there is also a contribution (typically dominant) from the spatial
curvature of the string worldsheets.
This equation can be
recovered by extremizing an action for the
embedding of $\mc{C}_{q+1}$ in the blackfold worldvolume,
\eq\label{hatI}
\hat I=\int_{\mc C_{q+1}}d\hat V_{(q+1)}\,\Phi\,,
\eeq
where $d\hat V_{(q+1)}$ is the measure on $\mc C_{q+1}$ and the
variations deform the embedding of $\mc C_{q+1}$ inside $\mc W_{p+1}$.
Using our solution \eqref{Phisol}, \eqref{totpot}, we can write this as
the action for a point particle of effective mass $\Phi_H$ that moves
along the worldvolume directions transverse the current,
\eq
\hat I=\Phi_H\int d\tau
\eeq
where $\tau$ is the proper time along the trajectory of the ``particle''.

\subsection{Extremal limit at the blackfold boundary}
\label{subsec:extremalbdry}

For worldvolumes with boundaries the presence of charges (or dipoles)
introduces a novelty compared to neutral blackfolds.

We have determined in \eqref{r0vanish} that $r_0$ must vanish at the
boundary of a blackfold, and for a stationary solution $r_0$ is given by
\eqref{sol1brane}. Then, for a neutral blackfold we must have
$\ga\to\infty$, \ie $k$ becomes null at the boundary, which may
happen because the blackfold meets a Killing horizon of the background,
or because the fluid velocity locally reaches the speed of light. In the
latter case, it must be noted that even if the lightlike-boost limit of
a \textit{uniform} neutral black brane results in a naked singular
solution, for a blackfold this is a phenomenon localized at the boundary
of the worldvolume and the full horizon of the blackfold can still be regular,
as is the case with the disks of \cite{Emparan:2009cs,Emparan:2009at}
and those discussed below.

When charges are present $r_0$ can vanish not only where $k$ becomes
a null vector, but also at other boundaries where, for a given $\Phi_H$,
\beq
|\hat h|^{1/2}=\frac1{(2\pi)^q}\frac{\Phi_H}{\sqrt{N}}\,.
\eeq
Then $k$ can remain timelike, but the charge boost $\alpha$ diverges:
the black brane becomes (locally) \textit{extremal}. Instances of this
phenomenon will appear in secs.~\ref{sectiondisk0} and \ref{subsec:annulus}.

When $T$ is finite and non-zero this extremal limit will occur only in a
localized manner, at some of the boundaries of the blackfold but not
away from them. Therefore even if the extremal limit of the uniform
brane may be singular, this need not imply a singularity of the
blackfold horizon and indeed in secs.~\ref{sectiondisk0} and
\ref{subsec:annulus} we present evidence of this.

\subsection{Extrinsic equations and action for stationary blackfolds}

We make the natural assumption that the vector fields $k$ and $\psi$ of
the previous analysis are the pullbacks to the worldvolume of commuting
Killing vectors of the background spacetime. Thus equations
\eqref{nablau}, \eqref{acceleration} and \eqref{Dva} are also satisfied
as equations in the background.
The components of $\hat K^\mu$
in directions orthogonal to the worldvolume, which enter in the
extrinsic equations
\eqref{exteqn1}, can now be obtained from
\beq
\hat K^\mu=-\hat\perp^{\mu\nu}\p_\nu\ln\lp|\hat h|^{1/2}\rp\,
\eeq
with $|\hat h|^{1/2}$ as in \eqref{hathsol} and
$\hat\perp^{\mu\nu}=g^{\mu\nu}-\hat h^{\mu\nu}$.

Using the intrinsic worldvolume solutions to the stationarity conditions, the
extrinsic equations \eqref{exteqn1} reduce to
\eqa\label{extrinsic}
K^\rho%&=&-n\perp^{\rho\mu}\p_\mu\ln\lb\ga\lp\cosh\al\rp^{N}\rb\nonumber\\
&=&n\perp^{\rho\mu}\p_\mu\ln r_0\nonumber\\
&=&\perp^{\rho\mu}\p_\mu\ln(-P_\perp)\,,
\eeqa
both for $q=0,1$.

Applying the results of \cite{Emparan:2009at}, this equation can be
equivalently found by varying, under deformations of the brane
embedding, the action
\beq
I=-\int_{\mc{W}_{p+1}}\!\!\!d^{p+1}\si\sqrt{-h}\,P_\perp\,.
\label{stataction}
\eeq
The form of this action is a familiar one for $p$-branes (\eg for Dirac
branes, where the tension $-P$ is uniform), and for thermodynamic
systems in equilibrium (if we recall that $-P_\perp$ is equal to the
Gibbs free energy density). It is remarkable that the resulting action
depends explicitly only on the thickness $r_0$ of the blackfold (and the
embedding coordinates), regardless of the presence of charges and
anisotropy.

Assume now that the background spacetime has a timelike Killing
vector $\xi$, canonically normalized to generate unit time translations
at asymptotic
infinity, and whose norm on the worldvolume is
\beq
-\xi^2|_{\mc W_{p+1}}=R_0^2(\si)\,.
\eeq
Let us further assume that $\xi$ is hypersurface-orthogonal, so we can
foliate the blackfold in spacelike slices $\mc{B}_p$ normal to $\xi$.
The unit normal to $\mc{B}_p$ is
\eq\label{redshift}
n^a =\frac1{R_0}\xi^a\,.
\eeq
Thus $R_0$ measures the (gravitational) redshift between worldvolume time and
asymptotic time. When integrating over $\mc W_{p+1}$ we can split the
trivial integration over the Killing time generated by $\xi$, which
gives an overall factor $\beta$ of the time interval, so
\eq
I=\frac{\be\Omega_{(n+1)}}{16\pi G}\int_{\mc{B}_p}dV_{(p)}\,R_0r_0^n,
\label{action}\eeq
where $dV_{(p)}$ is the integration measure on $\mc{B}_p$.

%Therefore, we expect the action (\ref{action}) to govern the
%blackfold embeddings in all generality.

This result immediately implies the condition eq.~\eqref{r0vanish} at
the blackfold boundary $\p\mc{B}_{p}$: the variations of the action that
deform the boundary give $\delta_{\p\mc{B}_p} I\propto r_0$, so the
thickness $r_0$ must vanish at it.

%%%%%%%%%%%%%%%%%%%%%%%%%%%%%%%%%
%%%%%%%%%%%%%%%%%%%%%%%%%%%%%%%%%
\subsection{Physical properties of stationary blackfolds and first law}

Once the intrinsic fluid equations are solved we can proceed with the
computation of the resulting thermodynamic quantities of the blackfold.
Here we extend and improve the analyses in
\cite{Emparan:2009at,Emparan:2009vd}.

\paragraph{Mass, angular momentum, entropy and charge.}

Let the stationarity Killing vector $k^\mu$ be given by a linear
combination of orthogonal commuting Killing vectors of the background
spacetime,
\eq\label{kxichi}
k^\mu=\xi^\mu+\sum_i\Om_i\chi_i^\mu,
\eeq
where $\xi$ is the generator of time-translations of the background that
we introduced above, and $\chi_i$ are
generators of angular rotations in the background spacetime normalized
such that their orbits have periods $2\pi$. The angular velocities of
the blackfold along these directions are then given by $\Om_i$.
According to \eqref{uk}, \eqref{redshift} and \eqref{kxichi},
\eq
\ga=-\frac{n^a u_a}{R_0}\,.
\eeq
Since $-n^a u_a$ is the gamma-factor for relativistic time-dilation in the
fluid, we see that $\gamma$ accounts for the redshifts caused by both gravitational and
local Lorentz-boost effects.

The mass and angular momenta are given by the integrals of the
corresponding densities over the worldvolume section $\mathcal B_p$,
\eq
M=\int_{\mathcal B_p}dV_{(p)}\,T_{ab}n^a\xi^b\,,\qquad
J_i=-\int_{\mathcal B_p}dV_{(p)}\,T_{ab}n^a\chi_i^b\,.
\label{bfMJ}\eeq
The total entropy is deduced from the entropy current $s^a=s(\si)u^a$,
\eq
S=-\int_{\mathcal B_p}dV_{(p)}\,s_a n^a=
\int_{\mathcal B_p}dV_{(p)}\,R_0\ga s(\si)\,,
\label{bfS}\eeq
and similarly, the 0-brane charge is
\eq
Q=-\int_{\mathcal B_p}dV_{(p)}\,J_a n^a=
\int_{\mathcal B_p}dV_{(p)}\,R_0\ga \mc Q(\si)
\qquad (q=0)\,.
\label{bfQ0}
\eeq
The string charge is obtained
by integrating the charge
density over the directions $\mc B_{p-1}^\perp$ in $\mc B_p$ that are
orthogonal to the current,
\eq
Q=-\int_{\mc B_{p-1}^\perp}dV_{(p-1)}\,J_{ab} n^a m^b,
\label{bfQ1}\qquad (q=1)\eeq
where $m$ is the unit spatial vector in $\mc C_{2}$ that is orthogonal
to $n$. We can find that
\beq
v=\frac{m+(m_a u^a)u}{\sqrt{1+(m_a u^a)^2} }
\eeq
by demanding that it lies along $\mc C_{2}$ and satisfies \eqref{uv}.
Then
\beq\label{jmn}
J_{ab} n^a m^b=-R_0\ga_\perp \mc Q\,,
\eeq
where
\eq
\gamma_\perp =\frac{\ga}{\sqrt{1+(m_a u^a)^2}}
\eeq
can easily be seen to have the interpretation of the Lorentz contraction
factor due to fluid velocity in directions transverse to the string
current (times the gravitational redshift). When the rotation is
parallel to the strings, then $\gamma_\perp=1/R_0$. Observe that here we
do not require the existence of a second Killing vector $\psi$, but if
it exists, then $m=\psi/|\psi|$. 

Along the spatial directions of the current we have
\beq
d^q\sigma\,|\hat h|^{1/2}=\frac{1}{\gamma_\perp}d\hat V_{(q)}
\eeq
where $d\hat V_{(q)}$ is the integration measure on $\mc{C}_q$, along
the direction of $m$.
 Then the global potential
in \eqref{totpot} can be written
\eq
\Phi_H=\int_{\mc C_{q}} d\hat V_{(q)} \frac{\Phi(\sigma)}{\gamma_\perp}
\eeq 
where for $q=0$ we recover $\gamma_{\perp}=\gamma$ and there is no
integral to perform. This is valid for generic stationary blackfolds, including those
for which
the solution \eqref{vsol} does not apply. If it does, like in the
examples we consider later where $m=\psi/|\psi|$, then $\ga_\perp$
can be taken out of the integral. 

Thus
\beqa
\Phi_H Q&=&\int_{\mc C_{q}}d\hat V_{(q)}\int_{\mc B_{p-q}^\perp} dV_{(p-q)}\Phi R_0 \mc Q=
 \int_{\mathcal B_p} dV_{(p)}R_0\Phi \mc Q\,.
\eeqa
Observe that, while the
charge density $\mc Q$ has
dimensions that depend only on $n$ but not on $q$, and the local
chemical potential $\Phi$ is dimensionless for any $q$, the integrated
charge and potential have different dimensions for
$q=0$ and $q=1$,
\eq
[G Q]=(\mathrm{Length})^{D-3-q}\,,\qquad [\Phi_H]=(\mathrm{Length})^{q}\,
\eeq
($G$ is Newton's constant).
\paragraph{Thermodynamic relations and first law.}

Let us now express the action \eqref{action} in terms of these
quantities. Begin by writing \eqref{stataction} as
\eq
I=\beta\int_{\mathcal B_p} dV_{(p)} n^a k_a P_\perp
\eeq
(since $n_a k^a=-R_0$), and use the generic thermodynamic relations \eqref{pressdiff} and
\eqref{GDint} in the stress tensor \eqref{Tabanis} to find that
\beqa
k_a P_\perp&=&T_{ab}k^b+\mc T s k_a +\Phi \mc Q k_a\nonumber\\
&=& T_{ab}\xi^b+\sum_i\Omega_i T_{ab}\chi_i^b
+ T s_a +\Phi \mc Q k_a\,.
\eeqa
We have used \eqref{uk}, \eqref{constT} and \eqref{kxichi} here.
Contracting with $n^a$ and integrating over $\mathcal B_p$ we find
\beq\label{Ithermo}
I=\beta W=\beta\lp M-TS-\sum_i \Omega_i J_i -\Phi_H Q\rp\,,
\eeq
so we recover the identification between the action and the
thermodynamic grand canonical potential $W$. Note that we have only used the
intrinsic equilibrium solution and have not imposed the extrinsic
equations. Actually, this derivation of \eqref{Ithermo} applies to any
stationary charged fluid, since we have only used the generic relations
\eqref{pressdiff}, \eqref{GDint}, but not the specific equation of
state of the blackfold fluid.

In a sense, \eqref{Ithermo} is the
integrated version of the local thermodynamic equation \eqref{GDint}, so
we should expect to have two more relations, from \eqref{pressdiff} and
the equation of state
\eqref{eos}. It is easy to show from these that
\eq
-P_\perp =\frac1{n}\mc T s
\eeq
which upon integration, and with $\beta=1/T$, gives
\beq\label{ITS}
I=\frac1{n}S\,.
\eeq
The remaining relation can be obtained through
similar manipulations, in the form
\beq\label{presmarr}
(D-3)M-(D-2)\lp TS+\sum_i\Omega_i J_i\rp-(D-3-q)\Phi_H Q=\mc T_{\rm tot}\,,
\eeq
where
\eq
\mc T_{\text{tot}}=-\int_{\mc B_{p}}\!dV_{(p)}\, R_0\lp\ga^{ab}+n^an^b\rp T_{ab}
\eeq
is the total tensional energy, obtained by integrating the local tension
 over the blackfold volume \cite{Harmark:2004ch,Kastor:2007wr}.
%It is amusing to observe that we can write it as
%\eq
%\mc T_{\text{tot}}=\mc T_{\text{tot}}^{\xi}+2U_\mathrm{grav}\,,
%\eeq
%where
%\eq
%\mc T_{\text{tot}}^{\xi}=-\int_{\mc B_{p}}\!dV_{(p)}\, R_0\lp\ga^{ab}+\xi^a\xi^b\rp T_{ab}
%\eeq
%may be interpreted as the total tensional energy relative to
%asymptotic observers that follow orbits of $\xi$, and
%\beq
%U_\mathrm{grav}=\frac12\int_{\mc B_{p}}\!dV_{(p)}\,(R_0^2-
%1)T_{ab}n^a\xi^b\,,
%\eeq
%is the gravitational energy of the blackfold in the
%background: $R_0^2-1$, which is analogous to $-1-g_{tt}$, has for small
%redshifts the interpretation of twice the Newtonian potential, which
%acts on the mass density $T_{ab}n^a\xi^b$.

The Smarr relation for charged black holes,
\beq\label{smarr}
(D-3)M-(D-2)\lp TS+\sum_i\Omega_i J_i\rp-(D-3-q)\Phi_H Q=0\,
\eeq
must be recovered when one considers a Minkowski background, where
$R_0=1$, % (and hence $U_\mathrm{grav}=0$), 
and furthermore the extrinsic
equations for equilibrium are satisfied. Thus, extrinsic equilibrium in
Minkowski backgrounds must imply
\beq\label{zerotension}
\mc T_{\text{tot}}=0\,.
\eeq
In
simple instances where Carter's equations reduce to a single equation,
one can simply impose this condition to derive the solution to the
extrinsic equations.
If the tensional energy did not vanish, it would imply the presence of
sources of tension acting on the blackfold, \eg in the form of conical or
stronger singularities of the background space.
It should be interesting to derive \eqref{zerotension}
as a consequence of \eqref{extrinsic}. Solutions with $R_0\neq 1$, which
do not satisfy \eqref{zerotension}, can
also be interesting and are in fact studied, for neutral blackfolds,
in \cite{Caldarelli:2008pz,Armas:2010hz}.

Since the action \eqref{stataction} is stationary under variations of
the embedding, with
$T$, $\Omega_i$ and $\Phi_H$ held constant, we see that solutions to the
blackfold equations satisfy the first law (see \cite{Emparan:2004wy,Copsey:2005se})
\eq\label{bhfirstlaw}
dM=T dS+\sum_i\Om_idJ_i+\Phi_HdQ
\eeq
under variations of the embedding functions $X^\mu(\si)$. Observe that
the scaling dimensions of $M$, $S$, $J_i$ and $Q$ are consistent with
their numerical
coefficients in the Smarr relation.

Let us emphasize that, while the thermodynamic action \eqref{Ithermo},
Smarr relation \eqref{smarr} and first law \eqref{bhfirstlaw} are
exactly valid for all black holes with $q$-brane charges or dipoles,
eqs.~\eqref{ITS} and \eqref{presmarr} instead hold only to leading order
in the limit where the blackfold construction applies.

\paragraph{Scalings.} Let us assume that all length scales along $\mc
B_{p}$ are of the same order $\sim R$ and that there are no large
redshifts, of gravitational or Lorentz type, over most of the blackfold
(this is naturally satisfied since the redshifts become large only close to the
boundaries).
Then we have
\beqa
M&\sim& R^p r_0^n(1+\hat\nu \sinh^2\al)\,,\qquad J\sim R^{p+1}r_0^n\,,\nonumber\\
Q&\sim&R^{p-q}r_0^n\sinh\al\cosh\al\,,\qquad S\sim R^p\label{scalings}
r_0^{n+1}(\cosh\al)^N\,,
\eeqa
where $\hat\nu$ is a non-zero pure number, obtained from $n$ and $N$, which we shall
assume to be of order one.
They satisfy
\beq
S(M,J,\al)\sim J^{-\frac{p}n}M^{\frac{D-2}n}\frac{(\cosh\al)^N}{(1+\hat\nu
\sinh^2\al)^\frac{D-2}n}\,,
\eeq
\ie
\beq
S(M,J,\Phi_H)\sim J^{-\frac{p}n}M^{\frac{D-2}n}f(\Phi_H)\,,
\label{entropyscaling}\eeq
so the entropy scales with $J$ and $M$ in the same way as in
the neutral case, and is modified by only a factor of a function of the potential.

%%%%%%
\section{Regime of applicability of blackfolds. Extremal and
near-extremal limits}
\label{sec:range}

The blackfold approach is applicable whenever a black hole can be
approximated locally by a black brane. This requires a separation of
scales on the horizon, with one large length along the directions that
can be regarded as the worldvolume of the brane, and a small length
measuring the size of the horizon in directions transverse to the
worldvolume, or more generally the length over which the gravitational
field close to the brane in directions transverse to the worldvolume
differs appreciably from flatness. The latter scale is the one that gets
integrated out in the effective description and allows to treat the
brane as a probe in a background spacetime.

The long scale is what we have referred to above as $R$, while for the
small scale, charged blackfolds introduce two different radii: one is
the energy-density radius of the black brane,
\beq
r_\vep\sim r_0 (1+\hat\nu \sinh^2\alpha)^{1/n}\sim r_0(\cosh\alpha)^{2/n}
\eeq
and the other is the charge-density radius
\beq
r_{\mc Q}\sim r_0(\sinh\al\cosh\al)^{1/n}\,,
\eeq
(these are convenient, but other choices, like the horizon radius of the
$S^{n+1}$, could have been made). Since $r_{\mc Q}$ cannot be parametrically
larger than $r_\vep$, we choose the latter to define the length scale of
the region that is integrated out. Then the
approach is applicable when $r_\vep\ll R$, \ie when
\beq\label{rangeval}
\frac{r_0}{R}\ll (\cosh\al)^{-2/n}\,.
\eeq
In the presence of charge densities, $r_0$ is not the
horizon size. Instead, it can be regarded as a `non-extremality
parameter', which can be made very small while physical quantities like
mass and charge remain finite and approach the extremal bounds. 

Let us now introduce the mass-length and angular momentum-length as in
\cite{Emparan:2009at},
\beq
\ell_M\sim (GM)^{\frac1{D-3}}\,,\qquad \ell_J\sim \frac{J}{M}\,.
\eeq
According to \eqref{scalings},
\eq
\ell_J\sim \frac{R}{1+\hat\nu
\sinh^2\al}
\eeq
which implies that, in contrast to neutral blackfolds,  the angular
momentum-length does not always set the size of the
blackfold, $R$, since $\ell_J$ can be much smaller than $R$ when extremality is approached.
Furthermore,
\eq\label{lMlJ}
\lp\frac{\ell_M}{\ell_J}\rp^{D-3}\sim\lp\frac{r_0}{R}\rp^n (1+\hat\nu
\sinh^2\al)^{D-2}
\sim\lp\frac{r_0}{R}\rp^n (\cosh\al)^{2(D-2)}\,,
\eeq
which when combined with \eqref{rangeval} implies that the blackfold
approach requires
\beq\label{rangeval2}
\frac{\ell_M}{\ell_J}\ll \cosh^2\alpha\,.
\eeq
When the charge parameter $\alpha$ is moderate, this regime of validity
 is the ultraspinning regime
$\ell_M\ll\ell_J$.
However, when $\alpha$ is very large and therefore the blackfold is
close to having maximal (extremal) charge, this condition does not
impose any hierarchy between
$\ell_M$ and $\ell_J$ and in particular a large enough $\alpha$ may
allow $\ell_J\ll \ell_M$.

Thus there may exist regimes of slowly rotating black holes which can be
described as blackfolds, as long as they are sufficiently near extremality. However,
the converse statement that the rotation must be small (\ie
$\ell_J\lsim \ell_M$)
for any near-extremal blackfold, need not be true. As we will see now, it
is true for $q=0$ charged blackfolds, but it never
occurs for $q=1$ string-dipole blackfolds.

Using
\eqref{Ithermo}, \eqref{ITS} and
\eqref{smarr} we derive
\beq
M-(q+1)\Phi Q=\frac{D-2}n TS\,,
\eeq
and
\beq
\sum_i\Omega_i J_i -q\Phi_H Q=\frac{p}n TS\,.
\eeq
These relations are valid for black holes constructed as blackfolds in
backgrounds where $R_0=1$, \eg Minkowski backgrounds. Close to
extremality the term $TS$ is very small, so
\beq\label{nearext}
M\simeq (q+1)\Phi_H Q\,,\qquad \sum_i\Omega_i J_i \simeq q\,\Phi_H Q\,.
\eeq
Thus we see that for near-extremal charged black holes ($q=0$) the
rotation must be small. Since in this regime
$\Phi_H\simeq \sqrt{N}$, the mass approaches the BPS bound
$M=\sqrt{N}Q$ for the theories \eqref{theory}.
This approach to extremality can happen under two different
circumstances:
\begin{enumerate}

\item The vanishing of $\Omega J$ may signal a breakdown of the
blackfold approximation, in which not only the rotation, but indeed the
whole black hole horizon becomes small in the perturbative blackfold
expansion: $R$ becomes comparable to $r_\vep$ so the black
hole does not resemble a brane anymore. We expect that this horizon
becomes fairly rounded in all its dimensions, \ie qualitatively similar
to the extremal Kerr-Newman solution.

\item The blackfold approximation may instead remain valid, and the
worldvolume remain of finite size, while satisfying \eqref{rangeval}, but the
brane approaches extremality. In the limit, the fluid becomes a `charged
dust', with no pressure on the worldvolume. Static extremal black holes
of Einstein-Maxwell-dilaton theory satisfy a no-force condition that
allows to construct exact solutions, in the Majumdar-Papapetrou class,
using harmonic functions with arbitrary distributions of sources, in
particular continuous ones (all of which have singular horizons). When
these are distributed along a $p$-dimensional submanifold, we obtain a
brane of static charged dust. The blackfold approach describes how this
extremal singular brane can be `thermalized' into a near-extremal black
brane with a regular horizon, which requires only a small rotation to balance
the small tensions that appear. Thus, such black holes have horizons
which are much longer in some directions than in others without being
ultraspinning.
\end{enumerate}
These two possibilities occur in the
solutions that we construct in the next sections.

Let us now analyze blackfolds with $q=1$ string dipoles. In the
extremal limit we find
\beq\label{extvirial}
\sum_i\Omega_i J_i
=\Phi_H Q=\frac{M}2\,.
\eeq
In this case the rotation cannot vanish at extremality, since the string
energy $\Phi_H Q$ sets a lower limit to the total rotational energy
$\Omega J$. The reason is that, even if the strings may satisfy a
no-force condition in directions transverse to them, one still needs to
counterbalance their tension with centrifugal repulsion along directions
parallel to them.

Eq.~\eqref{extvirial} says that the energy of extremal black holes with
string dipole is equally distributed (`virialized') among the total
momentum $\Omega J$ and the energy in string tension $\Phi_HQ$. Such
relations had already been identified for a variety of constructions (in
supergravity and in string theory) of dipole rings with $N=1$ in
\cite{BlancoPillado:2007iz}. We have now shown that they hold for all
extremal black holes with string dipoles, in the ultraspinning limit
where the leading order blackfold approximation applies and backreaction
effects can be neglected. At the next order, the self-interaction of the
black brane modifies its energy and \eqref{extvirial} receives
corrections.

Note also that the relation \eqref{extvirial} is not a BPS bound.
Solutions with only dipoles, and no net charges, cannot be
supersymmetric.

\subsection{Extremal and near-extremal limits of known charged rotating
black holes} \label{subsec:extlim}

The existence of extremal and near-extremal black holes that admit a
description as blackfolds can be illustrated using known exact
solutions. To this effect we shall consider the
charged rotating black holes in Kaluza-Klein theory (\ie $N=1$) in
$D\geq 6$ as presented in \cite{Kunz:2006jd}, and the five-dimensional
charged rotating rings of the same theory in
\cite{Elvang:2003yy,Elvang:2003mj,Elvang:2004xi}. A few
other examples, \eg with $N=2$, could be analyzed but they do not add
any new features. The extremal limit of solutions with string dipoles
can be studied in the dipole black rings of \cite{Emparan:2004wy}, but
this analysis has been essentially done already in \cite{BlancoPillado:2007iz} so we
will omit it here.

Let us first study the extremal limit of the KK MP black holes. We refer
to appendix~\ref{app:KKbh} for the solution and the
definition of quantities in this section.
The extremal limit we are interested in is obtained by taking
\beq
\alpha\to\infty\,,\qquad m\to 0\,,\qquad m\sinh^2\alpha=\hat m~~\mathrm{fixed,}
\eeq
while the rotation parameters $a_i$ remain fixed.
In this limit the metric becomes
\beq\label{extball}
ds^2=- h^{-\frac{D-3}{D-2}}dt^2+h^{\frac1{D-2}}ds^2(\mathbb{R}^{D-1})
\eeq
where
\beq
ds^2(\mathbb{R}^{D-1})=F dr^2+\epsilon r^2
d\nu^2+(r^2+a_i^2)(d\mu_i^2+\mu_i^2d\phi_i^2)
\eeq
is the metric of the flat space $\mathbb{R}^{D-1}$ in spheroidal
coordinates, as can be seen by changing to coordinates
\beq
\rho_i=\sqrt{r^2+a_i^2}\,\mu_i\,,\qquad z= r\,\nu\,,
\eeq
in which
\beq\label{flatsp}
ds^2(\mathbb{R}^{D-
1})=\epsilon dz^2+d\rho_i^2+\rho_i^2d\phi_i^2\,.
\eeq
The function
\beq
h=1+\frac{\hat m r^{2-\epsilon}}{\Pi F}
\eeq
can readily be seen to be a harmonic function in $\mathbb{R}^{D-1}$ (see app.~\ref{app:KKbh}
for the definition of $\epsilon,\Pi,F$ etc.). Since also
the electric field and the dilaton become
\beq
A_t=\lp h^{-1}-1 \rp dt\,,\qquad \phi=-\frac14 a^\textsf{KK}\ln h\,.
\eeq
we see that the solution belongs in
the Majumdar-Papapetrou class in $D$ spacetime dimensions.

The conventional static extremal limit is recovered when all $a_i=0$, so
$h=1+\hat m/r^{D-3}$ is a harmonic function with a pole at $r=0$ where
the solution has a naked pointlike singularity.
We consider now a more general case where a number
\eq\label{srots}
s< \frac{D-3}{2}\,,
\eeq
of the
possible rotation parameters $a_j$, $j=1,\dots, s$ are non-zero,
while the others vanish, $a_k=0$,
$k=s+1,\dots,\left\lfloor\frac{D-1}{2}\right\rfloor$.
The condition \eqref{srots} means that if $D$ is even at least one of the rotation parameters
vanishes, and if $D$ is odd at least two of the rotations vanish.
Then $h$ has a pole at the locus $r=0$, but now this
is the $2s$-dimensional ball
\beq
0\leq \rho_j\leq a_j\,,\qquad j=1,\dots,s\,
\eeq
in \eqref{flatsp}.
In this form, we see that the extremal limit is a solution that can be
described as a distribution of charged extremal black holes smeared over
an even-ball. Since the forces among the extremal black holes cancel, there is no
tension on the ball and we can characterize it as a ball of `charged
dust'. Such extremal configurations have singular horizons of zero area
(so we incur in a slight misnomer by calling them black holes).

Let us now consider the solution close to the extremal limit, with
finite but small $m$. In this case, the coordinate radius for the
horizon, where $\Pi=mr_+^{2-\epsilon}$, can be approximately computed as
\beq
r_+\simeq \lp\frac{m}{\prod_{j=1}^s a_j^2}\rp^{1/(D-3-2s)}\,.
\eeq
The actual dimensions of the horizon are
\beq
\ell_\perp \simeq r_+(\cosh\alpha)^{1/(D-2-2s)}\,,\qquad
\ell_\|{}_j\simeq a_j\,,
\eeq
in directions orthogonal and parallel to the rotation planes, respectively. We
see that the black hole has the pancaked shape, $\ell_\perp\ll \ell_\|$,
characteristic of the regimes that can be described in the blackfold approach.

However, despite the similarities this is not an ultraspinning regime of the
kind studied in \cite{Emparan:2003sy}: the angular
velocities and angular momenta approach zero in the extremal limit,
\beq
\Omega_i\to \frac{1}{a_i\cosh\alpha}\to 0\,, \qquad
J_i\to \frac{\Omega_{(D-2)}}{8\pi G}\frac{\hat m a_i}{\cosh\alpha}\to
0\,.
\eeq
The solution remains of finite extent, but
the area and temperature vanish, while the mass and charge remain
finite and saturate the BPS bound, as they must for a solution in the
class \eqref{extball}.

Another example of this limit is provided by the
five-dimensional charged black rings of
\cite{Elvang:2003yy,Elvang:2003mj,Elvang:2004xi}. For simplicity we only
consider, again, the charged black ring with $N=1$, which is given in
appendix~\ref{app:chring}. The limit we are interested in is
\beq
\lambda\to 2\nu\to 0\,,\qquad \alpha\to\infty\,, \qquad
\lambda\sinh^2\alpha=\hat q~~\mathrm{fixed,}
\eeq
while $R$ remains finite,
in which the solution becomes
\beq\label{extring}
ds^2=-h^{-2/3}dt^2+h^{1/3}ds^2(\mathbb{R}^{4})
\eeq
where
\beq
ds^2(\mathbb{R}^{4})=\frac{R^2}{(x-y)^2}
\lp(y^2-1)d\psi^2+\frac{dy^2}{y^2-1}+\frac{dx^2}{1-x^2}+(1-x^2)d\phi^2\rp
\eeq
is the metric of the flat space $\mathbb{R}^{4}$ in `ring'
coordinates (see \eg \cite{Emparan:2006mm}). So the limiting extremal metric, and also
the electric field
and the dilaton, are of the same kind as in the previous example, now with
a harmonic function
\beq
h=1+\hat q(x-y)
\eeq
that has poles in a circle of radius $R$ along the direction $\psi$ in
$\mathbb{R}^{4}$. Thus this solution is interpreted as a static ring of charged
dust. Near the extremal limit, where $\lambda\simeq 2\nu$ are small and
$\alpha$ is large, the radius of the $S^2$ of the ring is
\beq
r_+\simeq R\, \hat q^{1/4}\,\nu^{3/4}\,,
\eeq
which is much smaller than the $S^1$ radius $R$. We have found again
that the solution becomes black brane(string)-like even if the rotation
is small.

%%%%%%%%%%%%%%%%%%%%%%%%%%%%%%%%%%%%%%%%%%%%%%%%%%%%%%%%%%%%%%%%%%%%%%%%%%%%%%
%%%%%%%%%%%%%%%%%%%%%%%%%%%%%%%%%%%%%%%%%%%%%%%%%%%%%%%%%%%%%%%%%%%%%%%%%%%%%%

%%%%%%%%%%%%%%%%%%%%%%%%%%%%%%%%%%%%%%%%%%%%%%%%%%%%%%%%%%%%%%%%%%%%%%%%%%%%%%
%%%%%%%%%%%%%%%%%%%%%%%%%%%%%%%%%%%%%%%%%%%%%%%%%%%%%%%%%%%%%%%%%%%%%%%%%%%%%%
\section{Black holes with electric charge}
\label{sec:0charge}

Using the formalism for stationary charged blackfold developed in
section \ref{sec:statbfs}, now we build compact stationary blackfolds carrying electric
charge in a Minkowski background. They describe electrically charged black
holes in an asymptotically flat spacetime.

%%%%%%%%%%%%%%%%%%%%%%%%%%%%%%%%%%%%%%%%%%%%%%%%%%%%%%%%%%%%%%%%%%%%%%%%%%%%%%
\subsection{Disk solution: rotating charged black hole of spherical
topology in $D\geq6$}
\label{sectiondisk0}

We start with a simple rotating two-dimensional disk embedded in
Minkowski spacetime. The horizon of the corresponding black hole will be
topologically a sphere $S^{n+1}$ fibered over the disk ${\mathcal B}_2$.
The sphere radius vanishes on the boundary of the disk, resulting in a
spherical $S^{n+3}$ horizon. Since the radius of the disk is much larger
than the radius of the internal sphere, the event horizon is not round,
but flattened along the blackfold directions. In the neutral case, it
was shown in \cite{Emparan:2009vd} that this configuration reproduces
the physics of a pancaked MP black hole in $D=n+5$ dimensions, with a
single ultraspin. Here we add electric charge to this system. For
generic dilaton coupling the solutions are new,
but in Kaluza-Klein theory (when $N=1$) we can compare with the known
exact solution and find exact agreement. In the non-dilatonic case, the system
will describe for the first time the electrically charged generalization
of MP black holes in pure Einstein-Maxwell theory in the regimes
described in sec.~\ref{sec:range}.

Let us consider Minkowki spacetime as background, for which $R_0=1$.
Carter's equations are trivially solved for a flat embedding, so we can
restrict the analysis to the blackfold plane, with polar coordinates
$(r,\phi)$. The metric is
\beq
ds^2=-dt^2+dr^2+r^2d\phi^2\,.
\label{mink3}\eeq
Stationarity implies the fluid is rigidly rotating, and its velocity is
\beq
u=\ga\lp\frac{\p}{\p t}+\Omega\frac{\p}{\p \phi}\rp\,,\qquad \ga=\frac{1}{\sqrt{1-\Omega^2 r^2}}\,.
\label{ugamma}\eeq
Then, the intrinsic blackfold equations are solved by \eqref{constT} and
\eqref{refPhi}, \ie
\eq
\mc T=\ga T,\qquad\Phi=\ga\Phi_H.
\label{TPhisol}\eeq
$T$ and $\Phi_H$ are the temperature and potential at the
origin $r=0$ where the blackfold fluid is at rest.
The thickness $r_0(r)$ and the charge
parameter $\al(r)$ are then obtained by solving \eqref{sol1brane},
\eq
r_0(r)=\frac n{4\pi T}\lp1-\Om^2r^2\rp^{\frac{1-N}2}\lp1-\Om^2r^2-\frac{\Phi_H^2}N\rp^{N/2},\qquad
\tanh\al(r)=\frac1{\sqrt N}\frac{\Phi_H}{\sqrt{1-\Om^2r^2}}.
\label{disk0sol}\eeq
Observe now that the boundary where the radius $r_0$ vanishes is at $r=r_{\rm max}$ given by
\eq
r_{\rm max}=\frac1\Om\sqrt{1-\frac{\Phi_H^2}N}\,.
\eeq
This sets the radius of the disk and, in contrast to uncharged disks,
occurs before the velocity $u$ becomes lightlike. Instead $r_{\rm max}$ is
the radius at which the brane becomes locally extremal, $\Phi(r_{\rm
max})=\sqrt{N}$. Thus the disk radius is reduced by the electric charge.
The upper bound on the potential, $\Phi_H=\sqrt{N}$, is saturated for
extremal blackfolds, for which $\al\rightarrow\infty$.

Using \eqref{disk0sol} it is
straightforward to compute the charges of the blackfold by performing
the integrals \eqref{bfMJ} and \eqref{bfQ0} over the disk. They can be
easily expressed in terms of hypergeometric functions, and they
factorize into the neutral blackfold result, which carries all the
$(T,\Om)$ dependence, times functions that depend only on the potential
$\Phi_H$ through the dimensionless radius $\Om r_{\rm max}$,
\eqa
&&M=\frac{\Om_{(n+1)}}{8G\Om^2}\frac{n+3}{n+2}\lp\frac n{4\pi T}\rp^nI_M(\Om r_{\rm max}),\qquad\quad
J=\frac{\Om_{(n+1)}}{4G\Om^3}\frac1{n+2}\lp\frac n{4\pi T}\rp^nI_J(\Om r_{\rm max}),\nonumber\\
&&S=\frac{\pi\Om_{(n+1)}}{2G\Om^2}\frac1{n+2}\lp\frac n{4\pi T}\rp^{n+1}I_S(\Om r_{\rm max}),\qquad
Q=\frac{\Om_{(n+1)}}{8G\Om^2}\lp\frac n{4\pi T}\rp^n\Phi_HI_Q(\Om r_{\rm max})\,,
\label{disk0charges}\eeqa
with
\eqa
&I_M(x)&=\frac{n+2}{n+3}\,x^{nN}
\lb
\lp1-x^2\rp{_2F_1}\lp1+\frac n2(N-1),1;1+\frac{nN}2;x^2\rp\nonumber\right.\\
&&\qquad\qquad\qquad\quad
+\frac{n+1}{nN+2}x^2{_2F_1}\lp1+\frac n2(N-1),1;2+\frac{nN}2;x^2\rp\nonumber\\
&&\qquad\qquad\qquad\qquad
\left.-\frac{2x^2}{(nN+4)(nN+2)}{_2F_1}\lp1+\frac n2(N-1),2;3+\frac{nN}2;x^2\rp\rb,
\eeqa
\eqa
&I_J(x)&=
\frac{n(n+2)}{(nN+4)(nN+2)}x^{nN+4}\,{_2F_1}\lp1+\frac n2(N-1),2;\frac{nN}2+3;x^2\rp
\nonumber\\
&&\qquad
+\frac{n+2}{nN+2}\lp1-x^2\rp x^{nN+2}\,{_2F_1}\lp1+\frac n2(N-1),2;\frac{nN}2+2;x^2\rp,
\eeqa
\eqa
&I_Q(x)&=\frac1Nx^{nN}\,{_2F_1}\lp\frac n2(N-1),1;\frac{nN}2+1;x^2\rp,
\eeqa
\eqa
&I_S(x)&=\frac{n+2}{nN+2}x^{nN+2}\,{}_2F_1\lp\frac n2(N-1),1;\frac{nN}2+2;x^2\rp.
\label{disk0int}\eeqa
In the uncharged limit we have $\Om r_{\rm max}=1$ and these functions
simplify to $I_M(1)=I_J(1)=I_Q(1)=I_S(1)=1$.
%%%%%%%%%
% KK disk
In Kaluza-Klein theory, for which $N=1$, the
integrals \eqref{disk0int} can be expressed in terms of elementary
functions.
We obtain
\eqa
&&M=\frac{\Om_{(n+1)}}{8G(n+2)\Om^2}\lp\frac n{4\pi T}\rp^n\lp1-\Phi_H^2\rp^{\frac n2}\lp n+3-\Phi_H^2\rp,\quad
J=\frac{\Om_{(n+1)}}{4G(n+2)\Om^3}\lp\frac n{4\pi T}\rp^n(1-\Phi_H^2)^{\frac n2+1}
\nonumber\\
&&S=\frac{\pi\Om_{(n+1)}}{2G(n+2)\Om^2}\lp\frac n{4\pi T}\rp^{n+1}\lp1-\Phi_H^2\rp^{\frac n2+1},\qquad
Q=\frac{\Om_{(n+1)}}{8G\Om^2}\lp\frac n{4\pi T}\rp^n\Phi_H(1-\Phi_H^2)^{\frac n2}\,,
\label{diskKK}\eeqa
In appendix \ref{app:KKbh} we show that this blackfold result agrees
precisely with the exact analytic form of the Kaluza-Klein black hole of
\cite{Kunz:2006jd} in
the ultraspinning regime. This is good evidence that the charged disk
construction, with its new kind of boundary where the disk velocity
remains timelike, results into non-singular horizons.
We emphasize that the charges
\eqref{disk0charges} describe for the first time $N\neq1$ electrically
charged black holes in the limits where our approximations apply.

In agreement with \eqref{entropyscaling}, the entropy scales with
mass and angular momentum with the same powers as in the neutral case.
Finally, the charge per unit mass of these black holes is bounded from above, and obeys the same inequality \eqref{QMratio} as static charged black holes,
\eq
\frac QM\leq\frac{1}{\sqrt N}.
\label{QMratio0}\eeq
As discussed in sec.~\ref{sec:range}, these bounds are saturated for generic
extremal charged blackfolds.

\paragraph{Extremal limit.} In the extremal limit $\Phi_H\to\sqrt{N}$ we
have $\Om r_{\rm max}\rightarrow 0$ and therefore if $\Omega$ remains
finite the blackfold disk appears to vanish. More appropriately, the
black hole horizon becomes of size $\sim r_{\mc Q}$ in all its directions,
which pushes it outside the applicability of the blackfold approximation. This
is an explicit illustration of the first of the two possibilities of extremal limits of
charged blackfolds discussed in
sec.~\ref{sec:range}.

The second possibility discussed in sec.~\ref{sec:range} is also
realized. We could have $\Omega\to 0$ as the extremal limit is
approached, and by appropriately tuning the rate at which $\Omega$
vanishes, the disk can have arbitrary radius $r_\mathrm{max}$. Thus we
approach the extremal solution through a family of near-extremal, slowly
rotating charged black holes that are flattened on the rotation plane.
In the extremal limit we find a disk of charged dust, which can be found
as an exact solution of the Einstein-Maxwell-dilaton theory with a
singular horizon. The horizon of the near-extremal solutions gets
pancaked without being ultraspinning. For Kaluza-Klein dilaton coupling
$N=1$ these are the solutions described in sec.~\ref{subsec:extlim}.

%%%%%%%%%%%%%%%%%%%%%%%%%%%%%%%%%%%%%%
\subsection{Even-ball solution: charged black holes with multiple ultraspins}

The charged disk solution can be easily generalized to general
even-dimensional balls, with ellipsoidal shape. We take $p=2k$ and work
in flat Minkowski background with metric ($i=1,\ldots,k=p/2$)
\beq
ds^{2}=-dt^{2}+\sum_{i=1}^{k}\left(dr_{i}^{2}+r_{i}^{2}d\phi_{i}^{2}\right).
\label{evenmetric}\eeq
The velocity of the rigidly rotating fluid describing the stationary blackfold is given by
\beq
u=\ga\lp\frac{\p}{\p t}+\sum_{i=1}^k\Omega_i\frac{\p}{\p \phi_i}\rp\,,\qquad \ga=\frac{1}{\sqrt{1-\sum_{i=1}^k\Omega_i^2 r_i^2}}\,.
\label{ugammaball}\eeq
Then, the intrinsic blackfold equations are again solved by the
correctly redshifted temperature $T$ and potential $\Phi_H$
\eqref{TPhisol} as for the disk solution.
The thickness $r_0(r)$ and the charge
parameter $\al(r)$ are
\eq
r_0(r_i)=\frac n{4\pi T}\lp1-\sum\Om_i^2r^2_i\rp^{\frac{1-N}2}\lp1-\sum\Om_i^2r_i^2-\frac{\Phi_H^2}N\rp^{\frac N2},\quad
\tanh\al(r_i)=\frac{\Phi_H/\sqrt N}{\sqrt{1-\sum\Om_i^2r_i^2}}.
\label{ball0sol}\eeq
The thickness of the brane vanishes on the boundary of the ellipsoid $\mc E$ defined by
\beq
\sum_{i=1}^{k}\Omega_{i}^{2}r_{i}^{2}\leq x_m^2,\qquad
x_{m}=\sqrt{1-\frac{\Phi_{H}^{2}}N},
\eeq
From this we can also see that we have to require that $\Phi_{H}^{2}\leq N$.
We can now calculate the charges by performing the integrals
(\ref{bfMJ})-(\ref{bfQ0}) on $\mc E$. The dependences on $T$,
$\Omega_{i}$ and $\Phi_H$ factorize and we find that the charges are
given by the uncharged results times a potential dependent term,
\beqa
&&M=\frac{(2\pi)^{\frac{p}{2}}\Omega_{(n+1)}}{16\pi G\prod\Omega_{i}^{2}}\left(\frac{n}{4\pi T}\right)^{n}I_{M}(x_m),\qquad
J_{j}=\frac{(2\pi)^{\frac{p}{2}}n\Omega_{(n+1)}}{16\pi G\Omega_{j}\prod\Omega_{i}^{2}}\left(\frac{n}{4\pi T}\right)^{n}I_{J_{j}}(x_m),\\
&&S=\frac{(2\pi)^{\frac{p}{2}}\Omega_{(n+1)}}{4G\prod\Omega_{i}^{2}}\left(\frac{n}{4\pi T}\right)^{n+1}I_{S}(x_m),\qquad
Q=\frac{n(2\pi)^{\frac{p}{2}}\Omega_{(n+1)}}{16\pi G\prod\Omega_{i}^{2}}\left(\frac{n}{4\pi T}\right)^{n} \Phi_{h}I_{Q}(x_m),\nonumber
\eeqa
where the electric potential enters through the functions
\beqa
&&I_{M}(x_{m})=\int_{{\bf x}^2\leq x_{m}^{2}} \lp\prod_{i}x_{i}dx_{i}\right)\left(1-{\bf x}^{2}\right)^{\frac{n}{2}(1-N)-1}\left(x_{m}^{2}-{\bf x}^2\right)^{nN/2}
\left[n+1-{\bf x}^2+\frac{nN\lp1- x_m^{2}\rp}{x_m^2-{\bf x}^2}\right],\nonumber \\
&&I_{J_{j}}(x_{m})=\int_{{\bf x}^2\leq x_m^{2}}\lp\prod_ix_{i}dx_{i}\rp x_{j}^{2}\left(1-{\bf x}^2\right)^{\frac{n}{2}(1-N)-1}\left(x_m^{2}-{\bf x}^2\right)^{nN/2}\left[1+N\frac{1- x_m^{2}}{x_m^{2}-{\bf x}^2}\right],
\nonumber\\
&&I_{Q}(x_{m})=\int_{{\bf x}^2\leq x_m^{2}} \lp\prod_{i}x_{i}dx_{i}\right)\left(1-{\bf x}^2\right)^{\frac{n}{2}(1-N)}\left(x_m^{2}-{\bf x}^2\right)^{nN/2-1},
\nonumber\\
&&I_{S}(x_{m})=\int_{{\bf x}^2\leq x_m^{2}} \lp\prod_{i}x_{i}dx_{i}\right)\left(1-{\bf x}^2\right)^{\frac{n}{2}(1-N)}\left(x_m^{2}-{\bf x}^2\right)^{nN/2}.
\eeqa
We used ${\bf x}^2$ as a shorthand for $\sum x_i^2$.
With $p=2$, these expression reproduce the results for the rotating disk
of section \ref{sectiondisk0}, and in the uncharged ($\Phi_{H}=0$) limit
we recover the neutral even-ball quantities of \cite{Emparan:2009vd}.

%%%%%%%%%%%%%%%%%%%%%%%%%%%%%%%%%%%%%%%%%%%%%%%%%%%%%%%%%%%%%%%%%%%

%%%%%%%%%%%%%%%%%%%%%%%%%%%%%%%%%%%%%%%%%%%%%%%%%%%%%%%%%%%%%%%%%%%%%%%%%%%%%%
%%%%%%%%%%%%%%%%%%%%%%%%%%%%%%%%%%%%%%%%%%%%%%%%%%%%%%%%%%%%%%%%%%%%%%%%%%%%%%
\subsection{Odd-sphere solution with electric charge}
\label{subsec:oddsph0}

A natural solution to seek is the thin charged black ring.
As shown in \cite{Emparan:2009cs,Emparan:2009vd}, the blackfold construction of the
ring solution is the first occurrence of a larger family of blackfold
solutions whose worldvolumes wrap an odd-dimensional sphere $S^p$, with
$p=2k+1$. In general, with non-equal rotations $\Om_i$ on the Cartan
planes of the sphere, one has to solve a complicated differential
equation for the blackfold embedding. However, for equal rotations $\Om$
on all planes, the $S^p$ is a round sphere of radius $R$ and the
resulting solution is simple. We will therefore restrict to this simple
case in the analysis of odd-sphere blackfolds, and regard the black ring
as a particular case that can be recovered for $p=1$ as there is no
added complexity in considering a general $p$.

We embed the $S^p$ sphere in $\mathbb R^{p+1}$ with metric
\eq
dr^2+r^2\sum_{i=1}^{k+1}\lp d\mu_i^2+\mu_i^2d\phi_i^2\rp,\qquad
\sum_{i=1}^{k+1}\mu_i^2=1,
\label{oddmetric}\eeq
as a constant $r=R$ hypersurface. Then, the sphere is parameterized by $k+1$ Cartan angles $\phi_i$ and $k$ independent director cosines $\mu_i$. The velocity of a fluid rigidly rotating with equal angular velocity $\Om$ in all $k+1$ Cartan planes is given by
\eq
u=\ga\lp\frac\p{\p t}+\Om\sum_{i=1}^{k+1}\frac\p{\p\phi_i}\rp,\qquad
\ga=\frac1{\sqrt{1-\Om^2R^2}}.
\eeq
The extrinsic curvature of the sphere is easily seen to be $K^r=-p/R$
and then the extrinsic equation \eqref{exteqn0} is solved by choosing
its radius to satisfy
\eq
R=\frac1\Om\sqrt{\frac{p}{n+p+nN\sinh^2\al}}.
\label{R0odd}\eeq

This is the equilibrium radius for which the centrifugal force due to
the rotation of the blackfold balances the tension of the sphere.
In terms of the rapidity $\eta$, with $\tanh\eta=\Omega R$, \eqref{R0odd} is
\beq\label{equilsph}
\sinh^2\eta=\frac{p}{n(1+N\sinh^2\al)}\,,
\eeq
and we see that the local velocity needed to support the blackfold is
reduced by the presence of charge, as anticipated in sec.~\ref{subsec:bfstress}.
The fluid velocity decreases from the neutral limit $\Om R=\sqrt{p/(n+p)}$ to
a configuration with $\Om R=0$ in the extremal limit
$\al\rightarrow\infty$, $\Phi_H\to \sqrt{N}$. We will return to this
limit at the end of our analysis of the solutions.

The intrinsic equations are solved by \eqref{TPhisol}. The temperature $T$ and the potential $\Phi_H$ are constant over the sphere. Using \eqref{R0odd}, the latter can be expressed as
\eq
\Phi_H=\sqrt{nN}\sqrt{\frac{1+N\sinh^2\al}{n+p+n N\sinh^2\al}}\tanh\al.
\label{phi0odd}\eeq
To find the explicit dependence of the sphere radius $R(\Om,\Phi_H)$ on the angular velocity and electric potential, one must solve equation \eqref{phi0odd} for $\alpha$ and substitute the result in \eqref{R0odd} leading to
\beq
 R = \frac{\sqrt{n+2p+(nN-n-p)\frac{\Phi_H^2}{N}-\sqrt{\lp n-(nN+n+p)\frac{\Phi_H^2}{N}\rp^2+4n(n+p)\Phi_H^2\lp1-\frac{\Phi_H^2}{N}\rp}}}{\Om\sqrt{2(n+p)}}.
\label{radiusodd}\eeq
For simplicity, we choose to express all quantities in terms of $\al$ instead of $\Phi_H$.
Then, the constant thickness of the blackfold in terms of $T$ is
\eq
r_0=\frac{n^{3/2}}{4\pi T}\lp
\frac{1+N\sinh^2\al}{n+p+n N\sinh^2\al}\rp^{1/2}\lp\cosh\al\rp^{-N}\,.
\eeq
Since the worldvolume fields are constant, it is straightforward to
compute the thermodynamic variables for these blackfolds. The integrals
over the blackfold worldvolume reduce to a simple
multiplication by the size $R^p\Omega_{(p)}$ of the
sphere $S^p$.
Then
\eqa
&&M=\frac{\Omega_{(n+1)}\Omega_{(p)}}{16\pi G}
\frac{p^{p/2}}{\Omega^p}
\lp\frac{n^{3/2}}{4\pi T}\rp^n
\frac{\lp 1+N\sinh^2\al\rp^{n/2}\lp n+p+1+nN\sinh^2\al\rp}{\lp n+p+nN\sinh^2\al\rp^{(n+p)/2}\lp\cosh\al\rp^{nN}},
\label{ebrM}
\\
&&J=\frac{\Omega_{(n+1)}\Omega_{(p)}}{16\pi G}
\frac{p^{1+p/2}}{\Omega^{p+1}}
\lp\frac{n^{3/2}}{4\pi T}\rp^n
\frac{\lp 1+N\sinh^2\al\rp^{n/2}}{\lp n+p+nN\sinh^2\al\rp^{\frac{n+p}2}\lp\cosh\al\rp^{nN}},
\\
&&S=\frac{\Omega_{(n+1)}\Omega_{(p)}}{4G}
\frac{p^{p/2}}{\Omega^p}
\lp\frac{n^{3/2}}{4\pi T}\rp^{n+1}
\frac{\lp 1+N\sinh^2\al\rp^{n/2}}{n^{1/2}\lp n+p+nN\sinh^2\al\rp^{(n+p)/2}\lp\cosh\al\rp^{nN}},
\\
&&Q=\frac{\Omega_{(n+1)}\Omega_{(p)}}{16\pi G}
\frac{p^{p/2}}{\Omega^{p}}
\lp\frac{n^{3/2}}{4\pi T}\rp^n
\frac{\sqrt{nN}\sinh\al\lp1+N\sinh^2\al\rp^{(n-1)/2}}{\lp n+p+nN\sinh^2\al\rp^{(n+p-1)/2}\lp\cosh\al\rp^{nN-1}}\,,
\label{ebrQ}
\eeqa
Here $J$ is the momentum conjugate to $\sum_{i=1}^{k+1}\p_{\phi_i}$, and $\al=\al(\Phi_H)$ is given implicitly by (\ref{phi0odd}).

In can be checked that the entropy of these black holes scales with the mass and angular momentum exactly like their neutral counterparts, as predicted by \eqref{entropyscaling}.
Moreover, their charge to mass ratio satisfies precisely the same bound
\eqref{QMratio0} as the disk solutions and static charged black holes.

As an example in which the formulas do not get too cumbersome, we can
solve in the KK case ($N=1$) the relation between the parameter $\al$
and the potential $\Phi_H$, to obtain
\eq
\tanh\al=\Phi_H\sqrt{\frac{n+p}{n+p\Phi_H^2}}\,.
\eeq
In this case, the electric potential is bounded according to
$\Phi_H\leq1$ and the charges can be written explicitly in terms of the
potentials,
\eqa
&&M=\frac{\Omega_{(n+1)}\Om_{(p)}}{16\pi G}\frac{p^{p/2}}{\Om^p}\lp\frac{n^{3/2}}{4\pi T}\rp^n
\lp\frac{1-\Phi_H^2}{n+p}\rp^{\frac{n+p}2}
\frac{n+p+1-\Phi_H^2}{1-\Phi_H^2},\\
&&J=\frac{\Omega_{(n+1)}\Om_{(p)}}{16\pi G}\frac{p^{\frac p2+1}}{\Om^{p+1}}\lp\frac{n^{3/2}}{4\pi T}\rp^n\lp\frac{1-\Phi_H^2}{n+p}\rp^{\frac{n+p}2},\\
&&S=\frac{\Omega_{(n+1)}\Om_{(p)}}{4G}\frac{p^{p/2}}{\sqrt n\Om^p}\lp\frac{n^{3/2}}{4\pi T}\rp^{n+1}\lp\frac{1-\Phi_H^2}{n+p}\rp^{\frac{n+p}2},\\
&&Q=\frac{\Omega_{(n+1)}\Om_{(p)}}{16\pi G}\frac{p^{p/2}}{\Om^p}\Phi_H\lp\frac{n^{3/2}}{4\pi T}\rp^n
\lp\frac{1-\Phi_H^2}{n+p}\rp^{\frac{n+p}2-1}\,.
\eeqa
We stress that all these results describe electrically charged black
rings in the ultraspinning or near-extremal regimes when $p=1$. In the $N=1$ case, an
exact solution of this electric black ring exists, and agrees with our
results in the ultraspinning limit. We show this in
appendix~\ref{app:KKbh}. The near-extremal case was considered in
sec.~\ref{sec:range}.

\paragraph{Extremal limit.} We have found above that $\Omega R\to 0$ in the extremal
limit of these solutions. It is easy to see that also
$\Omega J\to 0$, which is in agreement with our general result in
sec.~\ref{sec:range} for $q=0$. The two possible realizations of
this limit that we discussed there, also occur in this example. The first one
results if we keep $\Omega$ finite, so $R$ must vanish. This really
means that the size of the blackfold sphere $S^p$ becomes too small, on
the scale of $r_{\mc Q}$, and thus beyond the leading order of the blackfold
perturbative construction. In this sense, this limit is similar to what
we have found for the extremal limit of disks keeping $\Omega$ finite,
only now with several equal spins instead of just one.

The second interpretation obtains if we keep $R$ finite, so now the
solution is not small, but is static. It is in fact a sphere of charged
dust, and the exact solution of Einstein-Maxwell-dilaton theory that it
corresponds to can be obtained in the Majumdar-Papapetrou form with
harmonic functions sourced on $S^p$, \ie it is a spherical smeared
distribution of extremal static black holes. Its horizon is singular,
but it can be regarded as a `good' singularity since by making it
non-extremal, which requires a small rotation, it becomes regular.

%%%%
\subsection{General products of odd-spheres}
\label{subsec:prodsph}

In \cite{Emparan:2009vd}
it was shown that the blackfold worldvolume can wrap any product of
odd-spheres, obtaining black holes in $D$-dimensional flat space with
horizon topology
\beq
\left(\prod_{p_{a}=\text{odd}}S^{p_{a}}\right)\times S^{n+1},\qquad \sum_{a=1}^{l}p_{a}=p.
\eeq
We now obtain their charged generalization. Following the notation of \cite{Emparan:2009vd},
we embed the blackfold in Minkowski spacetime with metric
\eq
dr_a^2+r_a^2\sum_{a,i}\lp d\mu_{a,i}^2+\mu_{a,i}^2d\phi_{a,i}^2\rp,\qquad
\sum_{i=1}^{k_a+1}\mu_{a,i}^2=1,
\label{prododdmetric}\eeq
where the $a$th sphere is parameterized by the director cosines and Cartan angles
$(\mu_{a,i},\phi_{a,i})$, in analogy to \eqref{oddmetric}.
Then, in a gauge where the spatial worldvolume coordinates are given by the $(\mu_{a,i},\phi_{a,i})$,
the embedding of the blackfold worldvolume $\mathcal{B}_{p}$ is
described by a collection of scalars $r_a=R_a(\mu_{1},\ldots,\mu_{k})$
that depend only on the director cosines, since we want to preserve the
rotational symmetry of the Cartan planes.
The induced worldvolume metric of the blackfold is then given by
\beq
ds^{2}_{p}=\sum_{a=1}^{l}\left[\sum_{i,j=1}^{k_{a}}\left[\left(\delta_{ij}+\frac{\mu_{a,i}\mu_{a,j}}{\mu_{a,k_{a}+1}^{2}}\right)R_{a}^{2}+\partial_{a,i}R_{a}\partial_{a,j}R_{a}\right]d\mu_{a,i}d\mu_{a,j} +R_{a}^{2}\sum_{i=1}^{k_{a}+1}\mu_{a,i}^{2}d\phi_{a,i}^{2}\right].
\eeq
The velocity field of a fluid rigidly rotating along the angles $\phi_{a,i}$ is,
\eq
u=\ga\lp\frac\p{\p t}+\sum_{a,i}\Om_{a,i}\frac\p{\p\phi_{a,i}}\rp,\qquad
\ga=\frac1{\sqrt{1-\sum_{a}R_a^2\sum_i\mu_{a,i}^2\Om_{a,i}^2}},
\eeq
and one can determine the set of equations that the embedding functions
$R_a(\mu_{a,i})$ must satisfy by writing down the worldvolume action
\eqref{action} and extremizing it. This yields a complicated
set of equations; the example for
a neutral $S^3$ blackfold can be found in \cite{Emparan:2009vd} and we
shall not attempt to do it here. 
%%%%
In the simpler case in which the scalars $R_a$ are constant defining the
radii of the $l$ spheres wrapped by the blackfold, and the angular
momenta of the $a$th sphere are all equal ($\Om_{a,i}=\Om_i$), the
equations become algebraic and can be easily solved to find the
equilibrium conditions
\eq
\Om_aR_a=\sqrt{\frac{p_a}{n+p+nN\sinh^2\alpha}}=\sqrt{\frac{p_a}p}\,{\Om R}\,,
\eeq
where $\Om R$ is the result, \eqref{R0odd} or \eqref{radiusodd}, for the equilibrium of a
single odd $p$-sphere blackfold.
%%%

With the shape of the blackfold at
hand, the intrinsic equations are solved as usual by \eqref{sol1brane},
and the thickness of the transverse $S^{n+1}$ is
\beq
r_{0}(\mu_{a,i})=\frac{n}{4\pi T}\lp1-\sum_{a}R_a^2\sum_i\mu_{a,i}^2\Om_{a,i}^2
\rp^{\frac{1-N}{2}}\left(1-\sum_{a}R_a^2\sum_i\mu_{a,i}^2\Om_{a,i}^2-\frac{\Phi_{H}^{2}}{N}\right)^{N/2}.
\eeq
The charges are finally easily found to be
\beqa
M&=&\frac{(2\pi)^{\frac{p+l}{2}}\Omega_{(n+1)}}{16\pi
G}\left(\frac{n}{4\pi T}\right)^{n}I_{M}(x_m,\Omega),\\
J_{b,j}&=&\frac{n(2\pi)^{\frac{p+l}{2}}\Omega_{(n+1)}\Omega_{b,j}}{16\pi G}\left(\frac{n}{4\pi T}\right)^{n}I_{J_{b,j}}(x_m,\Omega)\\
Q&=&\frac{n(2\pi)^{\frac{p+l}{2}}\Omega_{(n+1)}}{16\pi
G}\left(\frac{n}{4\pi T}\right)^{n} \Phi_{h}I_{Q}(x_m,\Omega),\\
S&=& \frac{(2\pi)^{\frac{p+l}{2}+1}\Omega_{(n+1)}}{8\pi G}\left(\frac{n}{4\pi T}\right)^{n+1}I_{S}(x_m,\Omega)
\eeqa
where the integrals are functions of the angular velocities, $\Om_{i}$, and the electric potential, given by
\beqa
\nonumber &&I_{M}(x_m,\Om)=\int_{\mathcal{\tilde{B}}} d\tilde{B}\left(1-\sum_{a}R_{a}^{2}\sum_{i=1}^{k_{a}+1}\mu_{a,i}^{2}\Om_{a,i}^{2}\right)^{\frac{n}{2}(1-N)-1} \left(x_m^{2}-\sum_{a}R_{a}^{2}\sum_{i=1}^{k_{a}+1}\mu_{a,i}^{2}\Om_{a,i}^{2}\right)^{\frac{nN}{2}}\\
&&
\qquad\qquad\qquad\qquad\qquad
\times
\lp n+1-\sum_{a}R_{a}^{2}\sum_{i=1}^{k_{a}+1}\mu_{a,i}^{2}\Om_{a,i}^{2}+\frac{nN\lp1- x_m^{2}\rp}{x_m^{2}-\sum_{a}R_{a}^{2}\sum_{i=1}^{k_{a}+1}\mu_{a,i}^{2}\Om_{a,i}^{2}}\rp,
\nonumber\\
&&I_{J_{b,j}}(x_m,\Om)=\int_{\mathcal{\tilde{B}}}d\tilde{B}R_{b}^{2}\mu_{b,j}^{2}\left(1-\sum_{a}R_{a}^{2}\sum_{i=1}^{k_{a}+1}\mu_{a,i}^{2}\Om_{a,i}^{2}\right)^{\frac{n}{2}(1-N)-1} \left(x_m^{2}-\sum_{a}R_{a}^{2}\sum_{i=1}^{k_{a}+1}\mu_{a,i}^{2}\Om_{a,i}^{2}\right)^{\frac{nN}{2}}
\nonumber\\
&&
\qquad\qquad\qquad\qquad\qquad\qquad\qquad\qquad
\times
\lp1+\frac{N\lp1- x_m^{2}\rp}{x_m^{2}-\sum_{a}R_{a}^{2}\sum_{i=1}^{k_{a}+1}\mu_{a,i}^{2}\Om_{a,i}^{2}}\rp,\nonumber\\
&&I_{Q}(x_m,\Om)=\int_{\mathcal{\tilde{B}}} d\tilde{B} \left(1-\sum_{a}R_{a}^{2}\sum_{i=1}^{k_{a}+1}\mu_{a,i}^{2}\Om_{a,i}^{2}\right)^{\frac{n}{2}(1-N)}
\left(x_m^{2}-\sum_{a}R_{a}^{2}\sum_{i=1}^{k_{a}+1}\mu_{a,i}^{2}\Om_{a,i}^{2}\right)^{\frac{nN}{2}-1},\nonumber\\
&&I_{S}(x_m,\Om)=\int_{\mathcal{\tilde{B}}} d\tilde{B} \left(1-\sum_{a}R_{a}^{2}\sum_{i=1}^{k_{a}+1}\mu_{a,i}^{2}\Om_{a,i}^{2}\right)^{\frac{n}{2}(1-N)} \left(x_m^{2}-\sum_{a}R_{a}^{2}\sum_{i=1}^{k_{a}+1}\mu_{a,i}^{2}\Om_{a,i}^{2}\right)^{\frac{nN}{2}},
\eeqa
with the integration measure defined as
\beq
d\tilde{B}= \prod_{a}R_{a}^{2k_{a}+1}\left(1+\mu_{a,k+1}^{2}\sum_{i=1}^{k_{a}}(\partial_{i}\ln R_{a})^{2}\right)^{1/2}\prod_{i=1}^{k_{a}}(\mu_{a,i}d\mu_{a,i}).
\eeq
Note that in all of these, whenever it appears,
$\mu_{a,k+1}^{2}=1-\sum_{i=1}^{k}\mu_{a,i}^{2}$. It is clear that this
reduces to the case of a single $S^{2k+1}$ sphere if we put $l=1$.

\bigskip

We conclude this section with a reminder that, as discussed in the
introduction, our approach does not capture the correct physics of
charged rotating black rings in the presence of Chern-Simons terms, such
as the solutions of minimal five-dimensional supergravity ($N=3$) in
\cite{Elvang:2004rt,Elvang:2004xi}. The appearance of Dirac-Misner
strings in the rotation form of the $S^2$
\cite{Elvang:2003mj,Elvang:2004xi} has the consequence that the
infinite-radius limit of the ring, as long as it is not a static string
of charged dust (as in sec.~\eqref{subsec:extlim}) but retains the
momentum, cannot have only 0-brane charge but also carries a string
charge, so it cannot be in the class of solutions we have studied. This
appears to be an effect entirely due to the Chern-Simons term, which is
absent in the five-dimensional theories with $N=1$ and $N=2$ for which
our construction of thin rings does work correctly. Our construction
above should also capture correctly the charged black rings
of the Einstein-Maxwell theory without a dilaton ($N=3$) and {\it
without} Chern-Simons term. Moreover, the possibility of Dirac-Misner
strings does not affect black holes with spherical horizon topologies
for which the $S^{n+1}$ is non-trivially fibered over the horizon.

%%%%%%%%%%%%%%%%%%%%%%%%%%%%%%%%%%%%%%%%%%%%%%%%%%%%%%%%%%%%%%%%%%%%%%%%%%%%%%
%%%%%%%%%%%%%%%%%%%%%%%%%%%%%%%%%%%%%%%%%%%%%%%%%%%%%%%%%%%%%%%%%%%%%%%%%%%%%%
\section{Black holes with string dipole}
\label{sec:1charge}

Now we apply our formalism to construct blackfolds with strings
dissolved in their compact worldvolume. They correspond to black holes
with string dipoles.

%%%%%%%%%%%%%%%%%
\subsection{Annulus solution: prolate black ring with string dipole in $D\geq 6$}
\label{subsec:annulus}

As in sec.~\ref{sectiondisk0}, we begin by looking for black 2-folds with worldvolume
spread in axially symmetric way in a geometry \eqref{mink3}, so the
velocity is again of the form \eqref{ugamma} and the temperature
determined according to \eqref{constT}. The unit
vector $v$ along the string
directions is
\beq
v=\frac{\ga}{r}\lp \frac{\p}{\p \phi}+\Omega r^2\frac{\p}{\p t}
\rp
\eeq
and the worldsheet area element
\beq
|\hat h|^{1/2}=r\,.
\eeq
The potential is then
\eq
\Phi=\frac{\Phi_H}{2\pi r}\,.
\eeq

Solving for $r_0(r)$ and $\alpha(r)$ we get
\eqa
r_0(r)&=&\frac{n}{4\pi T}\sqrt{1-\Om^2r^2}\;
\lp 1-\frac{\Phi_H^2}{N(2\pi r)^2}\rp^{N/2}\,,\\
\tanh\alpha(r)&=&\frac{\Phi_H}{\sqrt{N} 2\pi r}\,.
\eeqa
This fixes the solution. When computing the physical quantities we take
$n=\partial/\partial t$ and $m=r^{-1}\partial/\partial\phi$.

The worldvolume extends along values of $r$ between zeroes of $r_0$
\eq
r_{min}=\frac{\Phi_H}{2\pi\sqrt{N}}\leq r\leq
r_{max}=\frac{1}{\Om}\,.
\eeq
Note that both boundary conditions \eqref{stringbdry} and
\eqref{r0vanish} are satisfied by this solution. The upper bound, at
which $u$ becomes lightlike, is already familiar from disk blackfolds
and we expect the horizon to close smoothly there. However, the lower
bound is of a new type. Its origin can be understood in physical terms
from the Euler equation for the fluid,
\eq
\p_rP_\perp=\ga^2\lp r\om^2\lp\ep+P_\perp\rp-\frac{\mc Q\Phi}{r}\rp\,.
\eeq
The first term on the right is the usual centrifugal force. The second
term, proportional to $\mc Q\Phi$, has instead a centripetal effect on the
fluid that grows at smaller radii
like $1/r$. This is the force due to the strings, which is equal to
their tension times the extrinsic curvature, \ie $\mc Q\Phi/r$. Since
the centrifugal force decreases with $r$, and the stationary fluid must
rotate rigidly, it follows that at sufficiently small $r$ the tension
becomes too strong for the fluid to be able to remain in equilibrium.
This sets a minimum radius for the fluid, so the worldvolume is not a
disk but an annulus.

Since the size of the black brane's $S^{n+1}$ (not only $r_0$, but also
their area) vanishes at the inner and outer edges of the annulus, the
horizon of the black hole has the topology of a ring $S^1\times
S^{n+2}$, where the $S^{n+2}$ appears as the fibration of a sphere
$S^{n+1}$ of radius $r_0$ over an interval equal to the radial span
$r_{max}- r_{min}$ of the annulus.
Since the blackfold approximation
requires $r_{max}- r_{min}\gg r_0$ this horizon differs geometrically
from that of other black rings in that the $S^{n+2}$ is a prolate sphere,
highly elongated in the direction transverse to the strings. In contrast
to the flattening caused by centrifugal
forces, this elongation of the horizon can be regarded as caused by the
repulsion among parallel strings.
Therefore this is a qualitatively new class of dipole
black rings, which exist only in $D\geq 6$. The connection between the
two classes will be clarified below.

As discussed in sec.~\ref{subsec:extremalbdry}, at the boundary at the
inner rim where the bound on the potential is reached, the black brane
becomes extremal. Nevertheless it remains
non-extremal away from that edge.

While we expect from previous examples that the horizon remains
smooth at the outer edge, it is less obvious whether it will also remain
smooth at the inner edge. We found evidence that the 'extremal edge' was
not a problem for charged disks since they correctly reproduced
solutions that we know are smooth. In the present case there is no exact
solution that we can compare to, but still we will find evidence of good
behavior. At the very least, even if the solutions could have
a singularity at the inner edge, this would probably be a `good singularity',
as they appear to be connected to other ring solutions with good
singularities. Nevertheless, previous experience suggests that the edge
in the full solution is likely to be smooth.

The conserved charges are obtained by integrating the densities over the
blackfold annulus, and read
\eqa
&&M=\frac{\Omega_{(n+1)}}{8G\Om^2}\lp\frac{n}{4\pi T}\rp^n I_M\lp x_m\rp,\qquad
J=\frac{n \Om_{(n+1)}}{8G\Om^3}\lp\frac{n}{4\pi T}\rp^n I_J\lp x_m\rp,\nonumber\\
&&S=\frac{\pi\Om_{(n+1)}}{2G\Om^2}\lp\frac{n}{4\pi T}\rp^{n+1} I_S\lp x_m\rp,\qquad
Q=\frac{\Om_{(n+1)}}{32\pi^2 G}n\Phi_H\lp\frac{n}{4\pi T}\rp^n I_Q\lp x_m\rp,
\label{dbaCharges}\eeqa
where $x_m=\Omega r_{min}$ is the dimensionless inner
radius and the integrals are given by
\eqa
&I_M(x_m)&=\int_{x_m}^1\frac{(1-x^2)^{n/2}}{x^{nN-1}}\lp x^2-x_m^2\rp^{nN/2}
\lp1+\frac n{1-x^2}+\frac{nN x_m^2}{x^2-x_m^2}\rp dx\,,\nonumber\\
&I_J(x_m)&=\int_{x_m}^1\frac{(1-x^2)^{n/2-1}}{x^{nN-3}}\lp x^2-x_m^2\rp^{nN/2}dx\,,\nonumber\\
&I_Q(x_m)&=\int_{x_m}^1\frac{(1-x^2)^{n/2}}{x^{nN-1}}\lp x^2-x_m^2\rp^{nN/2-1}dx\,,\nonumber\\
&I_S(x_m)&=\int_{x_m}^1\frac{(1-x^2)^{n/2}}{x^{nN-1}}\lp x^2-x_m^2\rp^{nN/2}dx\,.
\label{dbaIntegrals}\eeqa
These integrals converge at the boundaries $x=1$, $x=x_m$.
They can be expressed in terms of hypergeometric functions,
but simplify to elementary functions in the
non-dilatonic, $D=6$ case,
for which $N=2$, $n=1$,
\eq
M=\frac{1}{6G\Om^2T}\lp1-x_m^2\rp^{3/2},\quad
J=\frac{1}{12G\Om^3T}\lp1-x_m^2\rp^{3/2},
\eeq
\eq
Q=\frac{\Phi_H}{32\pi^2GT}\lp\ln\frac{1+\sqrt{1-x_m^2}}{x_m}-\sqrt{1-x_m^2}\rp,
\eeq
\eq
S=\frac{1}{8G\Om^2T^2}\lb\frac13\lp1-x_m^2\rp^{3/2}-
x_m^2\lp\ln\frac{1+\sqrt{1-x_m^2}}{x_m}-\sqrt{1-x_m^2}\rp\rb.
\label{dbaChargesD6}
\eeq
Here $x_m=\Phi_H\Omega/(\sqrt{8}\pi)$.

Exactly like for the disk solution carrying 0-brane charge, the scaling
of the entropy with mass and angular momentum is unchanged with respect
to the neutral case and is given by (\ref{entropyscaling}). However, the
$\Om$-dependence of the charge has changed with respect to the
corresponding dependence for 0-brane charged disks, because we integrate
the density only on directions transverse to the strings carrying the
charge. As a consequence, $Q/M\propto\Om$, and there is no bound on the
charge.

\paragraph{Extremal limit.}
The annulus solutions exist only as long as
\eq
\Phi_H\leq \sqrt{N}\frac{2\pi}{\Omega}\,.
\eeq
When the upper bound here is saturated we have $r_{min}\to r_{max}$, so the
annulus becomes infinitely thin and the blackfold becomes both extremal
and lightlike. This suggests that this limit corresponds to an extremal
black ring with lightlike rotation and with zero temperature.
In general all the functions in \eqref{dbaCharges} smoothly go to zero as
$\Phi_H\rightarrow 2\pi\sqrt{N}/\Omega$ if we keep $T$ finite, but if instead we take
the limit as
\beq
T\to 0\,,\quad \Phi_H\rightarrow \sqrt{N}\frac{2\pi}{\Omega}\,,\quad
\mathrm{with}~~\frac{(2\pi\sqrt{N}-\Phi_H\Omega)^{(1+N)/2}}{T}~~\mathrm{finite}\,
\eeq
we can easily extract the dominant limiting terms in
\eqref{dbaIntegrals}. Taking into account that the ring radius is $R=1/\Omega$ we find that
\beq\label{JMQlim}
\sqrt{N}Q 2\pi R=\frac{J}{R}=\frac{M}{2}\,.
\eeq
These are the correct extremal limit relations derived in \eqref{extvirial}.
We regard this as evidence that the annulus blackfolds are physically
sensible solutions.

Note, however, that we should not expect that the entropy of the annulus
blackfold matches that of a dipole ring in this limit. The annulus is
locally equivalent to a (boosted) 2-brane with string charge, and the
latter is `smeared' along the direction transverse to the strings. Thus
the annulus can be regarded as a smeared distribution of concentric
rings. It is well-known that the extremal horizons do not behave
smoothly under smearing --- \eg if an extremal black string with regular
horizon is smeared in a transverse direction, the horizon of the
resulting string-charged 2-brane is singular. Indeed, the mismatch in
the entropies is even stronger in this case, since we are approaching
the extremal ring limit not from an extremal 2-brane annulus, but from
an annulus that is extremal only at its inner edge. This makes the
limiting entropy diverge if $N>1$. What we are seeing here is
that the approximations involved in this blackfold construction break
down when $r_{\mc Q}\sim r_{max}-r_{min}$. Nevertheless, the mass, angular
momentum and charge are measured at large distance from the brane and
are not sensitive to these horizon effects, hence the good behavior
shown in \eqref{JMQlim}.

%%%%%%%%%%%%%%%%%%%%%%%%%%%%%%%
\subsection{Solid ring and hollow ball solution: prolate black odd-spheres}

Given $p=2k$, a direct generalization of the annulus solution is given
by black $2k$-folds with worldvolume spread in axially symmetric way in
the geometry \eqref{evenmetric}. As in the electric even-ball case, the
velocity field of the fluid takes the form \eqref{ugammaball} and the
temperature is fixed by \eqref{constT}. To keep the worldvolume compact,
all angular velocities $\Om_i$ have to be non vanishing. Smearing the
charge-carrying strings on the planes of rotation, the unit vector $v$
along the string directions is determined by the Killing vector
\eq
\psi=\sum_{i=1}^k\varsigma_i\frac\p{\p\phi_i},
\eeq
through equations \eqref{zeta} and \eqref{vsol}. The constants $\varsigma_i$ determine 
the orientation, or polarization, of the strings in the worldvolume.
Then, according to
\eqref{hathsol}, the worldsheet area element is
\beq
|\hat h|^{1/2}=\sqrt{\ga^{-2}\sum_{i=1}^k\varsigma_i^2 r_i^2+\lp\sum_{i=1}^k\varsigma_i\Om_ir_i^2\rp^2}\,.
\eeq
The thickness $r_0(r_i)$ of the brane is given by \eqref{sol1brane}, and
the worldvolume of the brane fills the region $r_0(r_i)\geq0$, or
equivalently the region defined by the conditions
\eq
\sum_{i=1}^k\Om_i^2r_i^2\leq1\qquad\textrm{and}\qquad
\lp1-\sum_{i=1}^k\Om_i^2r_i^2\rp\sum_{i=1}^k\varsigma_i^2 r_i^2+
\lp\sum_{i=1}^k\varsigma_i\Om_ir_i^2\rp^2\geq\frac{\Phi_H^2}{(2\pi)^2N}.
\eeq
The first condition simply confines the blackfold worldvolume to an
ellipsoid with length of the axes set by the angular velocities $\Om_i$.
On the boundary of this ellipsoid, the velocity $u$ becomes lightlike.
The second constraint affects the blackfold shape only in presence of
string charges, drilling holes in the planes on which the polarization
of these strings lies. This leads to a variety of complicated
geometries. We content ourselves to illustrate it in the simple case of
$p=4$ blackfold, by studying two quintessential cases.

First, we align the strings to the rotation, by taking for example
$\Om_1=\Om_2=\Om$ and $\varsigma_1=\varsigma_2=1$. The blackfold
fills now a hollow ball, defined by
\eq
\frac{\Phi_H}{2\pi\sqrt N}\leq r\leq\frac1\Omega,
\eeq
where $r=\sqrt{r_1^2+r_2^2}$. This is a direct generalization of the
annulus blackfold obtained in the previous section. The black hole
event horizon has a $S^{n+1}$ fibered over the hollow ball $S^3\times
I$, whose radius vanishes on both exterior and interior boundaries (the
endpoints of the interval $I$), resulting in a topology $S^3\times
S^{n+2}$. This is an odd-sphere solution of a qualitatively different
kind than we shall find in the next section: although they
share the same topology, the $S^{n+2}$ is strongly elongated in the
radial direction of the hollow ball, and assumes a cigar-like shape,
with a longitudinal length (of $I$) much larger than the size of the
$S^{n+1}$,
\eq
\ell_\parallel\sim\frac1\Omega-\frac{\Phi_H}{2\pi\sqrt N}\gg\ell_\perp\sim r_0.
\eeq

This construction extends straightforwardly to any even $p$, leading to
prolate odd-sphere blackfolds that describe black holes with
$S^{2k-1}\times S^{n+2}$ horizon topology, the $S^{n+2}$ sphere being
prolate in the same sense as described above. By varying the
parameters $(\Om_i,\varsigma_i)$ one
can deform the blackfold sphere $S^{2k-1}$ into a spheroid.

The second representative solution that we exhibit has $\Omega_{1,2}\neq
0$ (and not necessarily equal) and the polarization
vector of the string dipole lying in the $(r_1,\phi_1)$ plane (\ie
$\varsigma_1=1,\varsigma_2=0$) so the strings and the rotation are not
fully aligned, and 
the fluid is restricted to the region
\eq\label{quadr}
\Om_1^2r_1^2+\Om_2^2r_2^2\leq1,\qquad
r_1^2\lp1-\Om_2^2r_2^2\rp\geq\frac{\Phi_H^2}{(2\pi)^2N}.
\eeq
The quadrant $r_1,r_2\geq 0$ represents the quotient space
$\mathbb{R}^4/U(1)^2$, where the two $U(1)$'s are the two rotations generated by $\partial_{\phi_{1,2}}$. In
this quadrant, as long as
\beq\label{phirange}
\Phi_H\leq\sqrt{N}\frac{2\pi}{\Omega_1},
\eeq
the region covered by \eqref{quadr} is non-vanishing, and the two bounding curves can be regarded as describing a lens (\ie the convex region bounded by two arcs), cut in half by the axis $r_2=0$ where
\beq
\frac{\Phi_H}{2\pi\sqrt{N}}\leq
r_1\leq \frac1{\Omega_1}.
\eeq
On the other hand $r_2$ varies between
\beq0\leq r_2\leq\frac1{\Omega_2}\sqrt{1-\frac{\Omega_1\Phi_H}{2\pi\sqrt{N}}}\,, 
\eeq
with the maximum being reached 
at
$r_1=\sqrt{\Phi_H/(2\pi\sqrt{N}\Omega_1)}$. The action of
$\partial_{\phi_2}$ on this half-lens turns it into a topological $B_3$
of lenticular shape.
The action of $\partial_{\phi_1}$ on this $B_3$ is free and gives a solid ring
$S^1\times B_3$, and this is the topology of the worldvolume. The
`transverse' sphere $S^{n+1}$ is fibered over this worldvolume,
shrinking to zero size at the ball boundaries, so the full horizon is
topologically $S^1\times S^{n+4}$, \ie a black ring, but one in which
three of the directions of the $S^{n+4}$, namely those along $r_1$ and
$(r_2,\phi_2)$, are much longer than the
others and can be comparable to the $S^1$ length along $\phi_1$. We call this
again a prolate black ring. Note the following
limits: when the inequality
\eqref{phirange} is saturated the lens
degenerates to the point $(r_1=1/\Omega_1$, $r_2=0)$, and we recover a
conventional dipole black ring; when $\Phi_H=0$ the solid ring becomes a
ball $B_4$ and we recover a neutral ultraspinning MP black hole; when
$\Omega_2\to 0$ we recover the annulus of the previous subsection times
the plane $(r_2,\phi_2)$; when $\Omega_1\to 0$ the strings and the
rotation lie on orthogonal planes: the rotation limits the maximum size
of the blackfold to $r_2\leq 1/\Omega_2$, while the strings put a
minimum to $r_1=\Phi_H/(2\pi\sqrt{N})$ on the plane $r_2=0$, but $r_1$ is
unbounded above so this does not describe a compact black hole.

By varying the parameters of the solutions, one can deform the hollow
ball into the solid ring solution. The calculation of the charges of
these solutions is straightforward and we will not undertake it, nor
will we embark in the study of the multiple shape these blackfolds can
assume for $p>4$ with generic parameters. We simply remark that the
construction in this section generalizes to blackfolds wrapping
the direct product of any odd-sphere and any odd-ball, as it can be seen
by taking equal rotation on all $k=p/2$ planes,
$\Om_1=\ldots=\Om_k=\Om$, and the polarization lying on the first $m$
planes, $\varsigma_1=\ldots=\varsigma_m=1$ and
$\varsigma_{m+1}=\ldots=\varsigma_k=0$. Defining
$\rho_1^2=r_1^2+\cdots+r_m^2$ and $\rho_2^2=r_{m+1}^2+\cdots+r_k^2$, the
positive quadrant in the $(\rho_1,\rho_2)$ plane represents the quotient
$\R^{p}/(SO(2m)\times SO(p-2m))$ and the fluid is again confined in a
half-lens-shaped region of this plane. Then, reasoning along the same
lines as above, it follows that the blackfold has topology
$S^{2m-1}\times B_{p-2m+1}$, and the corresponding black hole has a
$S^{2m-1}\times S^{n+p+2-2m}$ event horizon. The first factor is an
odd-sphere, but the solution differs from those we analyze next in that
the second spherical factor is prolate, much longer along $p-2m+1$
directions than along the others.

%%%%%%%%%%%%%%%%%%%%%%%%%%%%%%%%%%%%%%%%%%%%%%%%%%%%%%%%%%%%%%%%%%%%%%%%%%%%%%
%%%%%%%%%%%%%%%%%%%%%%%%%%%%%%%%%%%%%%%%%%%%%%%%%%%%%%%%%%%%%%%%%%%%%%%%%%%%%%
\subsection{Odd-sphere solution with string dipole}\label{sec:dipolesphere}

The blackfold equations can easily be solved for blackfolds wrapping
a round odd-sphere $S^p$, $p=2k+1$, of radius $R$ in a background Minkowski
spacetime\footnote{When $p=1$ these solutions are a particular case of
those considered in \cite{chargedbfolds}.}. The construction is very
similar to that in sec.~\ref{subsec:oddsph0}, with
the same choice for the velocity field $u$. The string polarization vector is
\beq
v=\frac{\ga}{R}\lp \sum_{i=1}^{k+1}\frac{\p}{\p \phi_i}+\Omega R^2\frac{\p}{\p t}
\rp
\eeq
and the worldsheet area element
\beq
|\hat h|^{1/2}=R\,.
\eeq
The potential is constant over the sphere
\eq
\Phi=\frac{\Phi_H}{2\pi R}\,,
\eeq
and so are $r_0$ and $\alpha$, too,
\eqa
r_0&=&\frac{n}{4\pi T}\sqrt{1-\Om^2R^2}\;
\lp 1-\frac{\Phi_H^2}{N(2\pi R)^2}\rp^{N/2}\,,\\
\tanh\alpha&=&\frac{\Phi_H}{\sqrt{N}2\pi R}\label{phidbr}\,.
\eeqa

The extrinsic curvatures $K^r=-p/R$ and $\hat K^r=-1/R$, give easily the
solution to the extrinsic equations
\eqref{exteqn1}
\eqa\label{Rsol1}
R=\frac1{\Omega}\sqrt{\frac{p+n N\sinh^2\al}{n+p+n N\sinh^2\al}}\,.
\eeqa

The product $\Om R$ is an increasing function of the charge (and the
potential), ranging from the neutral limit $\Om R=\sqrt{p/(n+p)}$ to
lightlike rotation $\Om R=1$ in the extremal limit
$\al\rightarrow\infty$, $\Phi_H\to \sqrt{N}$. This behavior is as
expected from our discussion at the end of sec.~\ref{subsec:bfstress}:
the addition of strings increases the tension so in a sphere of fixed
radius $R$ the angular velocity must be larger the more strings there
are. The expression \eqref{Rsol1} takes a simpler form in terms of the rapidity
$\eta$ such that $\tanh\eta=\Omega R$,
\beq\label{equileta}
n\sinh^2\eta=p+nN\sinh^2\alpha\,.
\eeq

Note however that in order to obtain the radius as function of $\Omega$
and $\Phi_H$ we must substitute \eqref{phidbr} in \eqref{Rsol1} and
solve, to find
\eq
R=\frac{\sqrt{
4\pi^2pN+(n+p-nN)\Omega^2\Phi_H^2+\sqrt{
\lp 4\pi^2pN-(n+p-nN)\Om^2\Phi_H^2\rp^2
+4n^2N^2\Omega^2\Phi_H^2}}}
{2\pi\Omega\sqrt{2N(n+p)}}\,.
\eeq
This result can also be obtained from the variation of the action.

We can find the thickness in terms of $T$, $\Omega$ and the
potential $\Phi_H$, or more conveniently of $\alpha$, as
\eq
r_0=\frac{n^{3/2}}{4\pi T}\lp n+p+n N\sinh^2\al\rp^{-1/2}\lp\cosh\al\rp^{-N}\,.
\eeq
Since the worldvolume fields are constant, it is straightforward to
perform the integrals and obtain
the thermodynamic variables for these blackfolds: $M, J, S$ involve simple
multiplication by the size $R^p\Omega_{(p)}$ of the
sphere $S^p$, while for $Q$ we have to multiply by the size of the subspace orthogonal to
the strings. Since these are aligned with the diagonal of the Cartan
subgroup of $S^{p}$, $p=2k+1$, this subspace is a $\mathbb{C}\mathrm{P}^{k}$ of radius
$R$
with size $R^{p-1}\Omega_{(p)}/(2\pi)$. Then
\eq
M=\frac{\Omega_{(n+1)}\Omega_{(p)}}{16\pi G}
\frac{1}{\Omega^p}
\lp\frac{n^{3/2}}{4\pi T}\rp^n
\frac{\lp p+nN\sinh^2\al\rp^{p/2}\lp n+p+1+2nN\sinh^2\al\rp}{\lp n+p+nN\sinh^2\al\rp^{(n+p)/2}\lp\cosh\al\rp^{nN}},
\label{dbrM}
\eeq
\eq
J=\frac{\Omega_{(n+1)}\Omega_{(p)}}{16\pi G}
\frac{1}{\Omega^{p+1}}
\lp\frac{n^{3/2}}{4\pi T}\rp^n
\frac{\lp p+nN\sinh^2\al\rp^{(p+2)/2}}{\lp n+p+nN\sinh^2\al\rp^{\frac{n+p}2}\lp\cosh\al\rp^{nN}},
\eeq
\eq
S=\frac{\Omega_{(n+1)}\Omega_{(p)}}{4G}
\frac{1}{\Omega^p}
\lp\frac{n^{3/2}}{4\pi T}\rp^{n+1}
\frac{\lp p+nN\sinh^2\al\rp^{p/2}}{n^{1/2}\lp n+p+nN\sinh^2\al\rp^{(n+p)/2}\lp\cosh\al\rp^{nN}},
\eeq
\eq
Q=\frac{\Omega_{(n+1)}\Omega_{(p)}}{32\pi^2 G}
\frac{1}{\Omega^{p-1}}
\lp\frac{n^{3/2}}{4\pi T}\rp^n
\frac{n\sqrt{N}\sinh\al\lp p+nN\sinh^2\al\rp^{(p-1)/2}}{\lp n+p+nN\sinh^2\al\rp^{(n+p-1)/2}\lp\cosh\al\rp^{nN-1}}\,,
\label{dbrQ}\eeq
with $\al$ given by the potential $\Phi_H$ in (\ref{phidbr}). Here $J$
is the angular momentum associated to $\sum_{i=1}^{k+1}\p_{\phi_i}$.

Again, we find that the entropy of these black holes scales with the
mass and angular momentum according to \eqref{entropyscaling}.
As for the annulus solution, the
ratio $Q/M$ is proportional to $\Om$ times a bounded function of
$\Phi_H$ only, and therefore there is no bound on the ratio of dipole
$Q$ to mass $M$. 

The horizon of these solutions is $S^{2k+1}\times
S^{n+1}$, and thus topologically the same as in the previous subsection.
However, here the last sphere factor is of much smaller size than the
first one in all directions. This small $S^{n+1}$ is geometrically round
to leading order in the blackfold approximation, but will be slightly
distorted at the next order. As before, the large $S^{2k+1}$ is exactly
round as long as all the rotations $\Omega_i$ are equal, but this can be
relaxed. 

When $n=1$, $p=1$, these solutions can be compared to the exact
five-dimensional dipole rings of \cite{Emparan:2004wy} in the limit of large angular
momentum. In appendix \ref{app:dipring} we show that there is perfect agreement.

\paragraph{Extremal limit.} This is obtained as $\alpha\to\infty$, \ie
$\Phi_H\to\sqrt{N}R$, so the rotation becomes lightlike,
\eq
R\Om\to 1\,.
\eeq
In order to keep the mass, spin and dipole charge finite we must also
keep
\eq
\frac{(\sinh\al)^{2-n(N+1)}}{T^{n}}\quad\mathrm{finite}\,,
\eeq
which means that if $n(N+1)>2$ then the temperature goes to zero. Since
$n\geq 1$, this condition will not be met only if $N<1$, which does not
seem to be relevant in string theory. The marginal case where the
temperature remains finite is $n=1$, $N=1$,
which includes the well-known case of five-dimensional string-like
objects with $N=1$ and their direct uplifts. The condition depends only on
the number of dimensions
transverse to the brane (given by $n$) and of $N$, and is independent of
$p$.

In this limit we recover the relations \eqref{extvirial} for $M$, $J$
and $Q$.
The entropy of the extremal black holes is finite and non-vanishing only
if $n(N-1)=2$, which can happen only when $(n=1,N=3)$, or $(n=2,N=2)$. These cases
are familiar when $p=1$, where they corresponds to extremal black
strings and rings in five and six dimensions, respectively. Other values of $p$ are
direct uplifts of them. In all these cases the entropy
is equal to
\beq
S=\frac{4\pi}{\sqrt{n(n+2)}}
\lp\frac{(n+2)\Omega_{(n+1)}\Omega_{(p)}}{16\pi G} \rp^{-1/n}
R^{-\frac pn}\lp\frac{M}{2}\rp^{\frac{n+1}n}
\eeq
As we discussed in the previous subsection, this
entropy does not match that of the extremal annulus solutions. Nevertheless
one can see that the curves of $S/M^{(n+1)/n}$ as function of $\Phi$
tend to cross around the region where the annulus entropy starts to
diverge.

\bigskip

Finally, it is straightforward to construct solutions with string
dipoles where the worldvolume is a product of odd-spheres. As they do
not add anything qualitatively new, we omit the details.

%%%%%%%%%%%%%%%%%%%%%%%%%%%%%%%%%%%%%%%%%%%%%%%%%%%%%%%%%%%%%%%%%%%%%%%%%%
%%%%%%%%%%%%%%%%%%%%%%%%%%%%%%%%%%%%%%%%%%%%%%%%%%%%%%%%%%%%%%%%%%%%%%%%%%
\section{Discussion and outlook}
\label{sec:outout}

We begin by discussing separately our results for each of the two types
of $q$-brane charge that we have considered in this paper, highlighting some
particular consequences and conjectures motivated by our study.

\paragraph{Charged rotating black holes.} We have constructed them in
theories with arbitrary dilaton coupling in any dimension $D\geq 5$. In
$D=5$ we have found them as black rings, and in $D\geq 6$ we have found
them with spherical topology, but also with more varied topologies, in
particular products of spheres. After the results of
\cite{Emparan:2009at,Emparan:2009vd} it may not be so surprising to find
this type of black holes when their rotation is very large: a
sufficiently large rotation can conceivably support the tension of an
extended brane with a compact worldvolume, and the addition of charge
should not change this qualitatively. Nevertheless, it is illustrative
of the power of the method that it has easily provided a window into the
elusive charged rotating black holes of Einstein-Maxwell(-dilaton)
theory with spherical topology, extending the Myers-Perry rotating black
holes to have charge. Their five-dimensional counterparts, although
expected to exist for any value of the dilaton, cannot become brane-like
and therefore fall outside the applicability of the blackfold
techniques. In five dimensions only string-like rings can be described
as blackfolds.

A bigger surprise has been to find another regime, not involving large
spins, in which the tension of the brane can be balanced: when the
charge is near extremality, the forces on the brane tend to cancel
leaving only a very small tension, and
therefore just a small rotation is needed to oppose it. We have found
that this can happen when the rotation occurs in a number $s$,
$1\leq s< (D-3)/2$, of all independent rotation planes. This is a regime in
which the black hole can become brane-like and therefore is amenable to
study as a blackfold.

There is a remarkable conclusion of this result. The static extremal solutions,
being charged dusts, are marginally stable solutions. However, the
addition of a small non-extremality, which gives them a non-singular
horizon, will make them unstable. This is because these
blackfolds are
locally black branes with 0-brane charge, and these are known to suffer
Gregory-Laflamme instabilities whenever they are non-extremal (more on
this below). In particular, this implies that in the
Einstein-Maxwell-dilaton theories
\eqref{theory} with $q=0$ and arbitrary $a$,
\begin{itemize}
\item Electrically charged black holes rotating in $s$ planes, with
$1\leq s< (D-3)/2$, become dynamically unstable when their charge is
sufficiently close to (but below) the extremal bound $Q=M/\sqrt{N} $.
This applies in particular to black holes with horizons of spherical
topology.
\end{itemize}
This may come as a surprise, since one would regard these
topologically-spherical rotating charged black holes as the natural
higher-dimensional counterparts of the Kerr-Newman solution, which is
expected to be stable, in particular at slow rotation. Moreover,
closeness to a BPS state might have suggested improved stability. What
changes these expectations in $D\geq 6$ is that rotation in $s\in
[1,(D-3)/2)$ planes close to extremality makes the black hole spread
along these planes and approach the geometry of an unstable
near-extremal black brane. This kind of instability will also be
present for charged black
holes with any other horizon topology, but since these are less
familiar, their instability may seem less unexpected. The case
where $s=(D-3)/2$ is not included in the statement above,
but in that case charged black holes with product-of-spheres topology
(secs.~\ref{subsec:oddsph0}, \ref{subsec:prodsph}), and in fact any that
are amenable to the blackfold approach, are expected to be
unstable close
to extremality since locally they resemble unstable thin black branes. These
include in particular the five-dimensional charged black rings with a single
angular momentum\footnote{As discussed above, the presence of
independent dipoles, or a Chern-Simons term, can take us away from the
regime where this conclusion applies.}.
The five-dimensional topologically-spherical charged black holes with
one spin do not become brane-like and cannot be constructed as
blackfolds, but if they get highly distorted near extremality
they could plausibly be unstable to $R$-mode deformations like in the
neutral case \cite{Emparan:2003sy,Shibata:2009ad}. 

Let us emphasize that this result is a conclusion, not a conjecture,
that follows from the perturbative blackfold construction of these black
holes. It allows us to establish that the instabilities will be eventually
present when the charge is sufficiently large, but not to determine the
precise parameter values for their onset. Following
\cite{Emparan:2003sy}, we expect this to occur before the near-extremal
brane-like regime is reached, and to be approximately marked by the
criterion that it is entropically favorable to split the black hole, in
a range of parameters after the appearance of a negative mode of the
thermodynamic Hessian \cite{Dias:2009iu,Dias:2010maa}. One expects also
the appearance of a new phase of black holes with pinches along the
horizon. It should be interesting to study this phenomenon in more
detail, in particular in the solutions with KK charge which are known in
closed form.

\paragraph{String-dipole black holes.} An important motivation
that we had to study these configurations was to try to find a black hole
with a horizon of spherical topology supporting a string dipole. It is
known for many of the theories in \eqref{theory} that a static black
hole cannot possess such dipole hair \cite{Emparan:2010ni}. One
might have hoped that a sufficiently large rotation could change this by
spreading the horizon along the plane of rotation, and therefore we
have attempted to construct the solution that could possibly give it,
namely a disk blackfold. However, such a
disk could be regarded as a concentric distribution of dipole
rings, and if these must rotate rigidly, then at a small enough radius the
centrifugal force will become too small to balance the tension of the
strings. This
forces the disk to open up a hole, so it
becomes an annulus.
The
black hole turns then into a black ring.

This suggests that a
horizon with spherical topology, no matter how large its rotation,
cannot support string dipole hair, nor, quite likely, any brane dipole
hair. We are therefore led to conjecture that
\begin{itemize}
\item Black hole horizons of spherical topology cannot support brane-dipole hair.
\end{itemize}
A still plausible but stronger conjecture is that
\begin{itemize}
\item Black hole horizons must have a non-contractible $q$-cycle in order to be
able to support $q$-brane dipole hair.
\end{itemize}
These conjectures refer to asymptotically
flat black holes of the theories \eqref{theory}. Other fields, or a
cosmological constant, might significantly alter the physics of the
system.

\bigskip

Like in the case of black holes built as neutral blackfolds, most of the
solutions in this paper are unstable to perturbations that create
ripples along the worldvolume \cite{Gregory:1994bj}. This
Gregory-Laflamme-type instability of objects that are locally like black
$p$-branes can be suppressed near extremality along directions parallel
to a $q$-brane current, but will persist in directions transverse to the
current \cite{Harmark:2007md}. Thus the only configurations in this
paper that are expected to be stable close to extremality are dipole
rings built as circular strings. The prolate dipole black rings will be
unstable, even close to extremality, to rippling along the elongated
directions. In general the instabilities should mark the bifurcation
into new phases with pinched horizons.

In this article we have only sampled solutions that illustrate the new
possibilities. Generalizations of other solutions in
\cite{Emparan:2009vd} to include electric charges and dipoles should be
straightforward. In particular, it is easy to find for both cases
helical black rings that have the minimal horizon symmetry allowed by
rigidity theorems.

%%%%%%%%%%%%%%%%%%%%%%%%%%%%%%%%%%%%%%%%%%%%%%%%%%%%%%%%%%%%%%%%%%%%%%%%%%
%%%%%%%%%%%%%%%%%%%%%%%%%%%%%%%%%%%%%%%%%%%%%%%%%%%%%%%%%%%%%%%%%%%%%%%%%%

\section*{Acknowledgments}

We are grateful to Troels Harmark, Vasilis Niarchos and Niels Obers for
discussions about blackfolds with charges.
MMC and BVP were supported in part by the FWO - Vlaanderen, project
G.0235.05 and in part by the Federal Office for Scientific, Technical
and Cultural Affairs through the ``Interuniversity Attraction Poles
Programme -- Belgian Science Policy'' P6/11-P. RE was partially
supported by DURSI 2009 SGR 168, MEC FPA 2007-66665-C02 and CPAN
CSD2007-00042 Consolider-Ingenio 2010. MMC was also supported by the ANR grant STR-COSMO,
ANR-09-BLAN-0157,
the ERC Advanced Grant  226371,
the ITN programme PITN-GA-2009-237920,
the IFCPAR CEFIPRA programme 4104-2, the ANR programme NT09-573739 ``string
cosmo'' and the PEPS-CNRS programme ``Cordes, evolution, anisotropies et
transitions''.

%%%%%%%%%%%%%%%%%%%%%%%%%%%%%%%%%%%%%%%%%%%%%%%%%%%%%%%%%%%%%%%%%%%%%%%%%%%%%%
%%%%%%%%%%%%%%%%%%%%%%%%%%%%%%%%%%%%%%%%%%%%%%%%%%%%%%%%%%%%%%%%%%%%%%%%%%%%%%
\appendix

%%%%%%%%%%%%%%%%%%%%%%%%%%%%%%%%%%%%%%%%%%%%%%%%%%%%%%%%%%%%%%%%%%%%%%%%%%%%%%
%%%%%%%%%%%%%%%%%%%%%%%%%%%%%%%%%%%%%%%%%%%%%%%%%%%%%%%%%%%%%%%%%%%%%%%%%%%%%%
\section{$p$-brane solutions with diluted electric $q$-brane charges}
\label{app:branesol}

In this appendix we build the generic black $p$-brane solution,
carrying $q$-brane charge and with horizon topology
$S^{n+1}\times\R^{p}$. We construct them starting from the charged,
spherically symmetric black holes of $d=n+3$ dimensional
Einstein-Maxwell dilaton (EMD) theory, proceeding in two steps. First, we
uplift these solution to $n+q+3$ dimensions, to obtain $q$-branes
charged under a $q+1$-form gauge potential isotropic in the $q$ extended
dimensions. In a second uplift, we add $p-q$ extra extended dimensions,
to obtain the desired solution.

More specifically, we start from EMD theory in $d$ dimensions with
dilaton coupling $\tilde a$ and action
\eq
I=\frac1{16\pi G}\int\!\! d^dx\,\sqrt{-\tilde g}
\left[\tilde R-2\lp\tilde\nabla\tilde\phi\rp^2-\frac{e^{-2\tilde a\tilde\phi}}{4}\tilde F^2\right].
\label{EMDaction}\eeq
Here, we are using the tilde to mark the $d$-dimensional quantities and
contrast them to the corresponding quantities of the uplifted solution. The Maxwell field strength is $\tilde F=\dd\tilde B_{[1]}$, with $\tilde B_{[1]}$ being the electromagnetic potential one-form.

The spherically symmetric, electrically charged black hole solutions of
this theory where obtained by Gibbons and Maeda in
\cite{Gibbons:1987ps}. Their metric can be put in the form
\eq
d\tilde s^2=-\frac{f}{h^{\tilde A}}dt^2+h^{\tilde B}\lp\frac{dr^2}{f}+r^2d\Omega_{n+1}^2\rp
\label{GMmetric}\eeq
with the functions $f(r)$, $g(r)$ and the coefficients $\tilde A$, $\tilde B$ given by
\eq
f(r)=1-\frac{r_0^n}{r^n}\,,\quad
h(r)=1+\frac{r_0^n}{r^n}\sinh^2\alpha\,,\quad
\tilde A=\frac{4n}{2n+(n+1)\tilde a^2}\,,\quad
\tilde B=\frac{4}{2n+(n+1)\tilde a^2}\,.
\eeq
This metric describes a spherical black hole, with an event horizon
located at $r=r_0$ and an electric charge determined by the parameter
$\al$. The corresponding dilaton field and Maxwell potential read
\eq
\tilde\phi(r)=-\frac{(n+1)\tilde a}{2n+(n+1)\tilde a^2}\,\ln h(r)\,,\qquad
\tilde B_{[1]}=-\tilde\Phi(r)\dd t\,,
\eeq
where we have defined the electric potential
\eq
\tilde\Phi(r)=2\sqrt{\frac{n+1}{2n+(n+1)\tilde a^2}}\,\frac{r_0^{n}}{r^nh(r)}\,\sinh\al\cosh\al.
\eeq
The mass and electric charge of this black hole are
\eq
M=\frac{\Omega_{(n+1)}}{16\pi G}r_0^n\lp n+1+N n\sinh^2\al\rp,
\qquad
Q=\frac{\Omega_{(n+1)}}{16\pi G}n\sqrt{N}r_0^n\sinh\al\cosh\al,
\eeq
and their ratio is bounded from above according to
\eq
\frac QM\leq\frac1{\sqrt N}.
\label{QMratio}\eeq
Extremal black holes saturate this bound.

%%%%%%%%%%%%%%%%%%%%%%%%%%%%%%%%%%%%%%%%%%%%%%%%%%%%%%%%%%%%%%%%%%%%%%%%
\subsection{Electrically charged $q$-branes}

Let us consider General Relativity in $D=d+q$ dimensions, coupled to a
dilaton field $\phi$ and a $(q+1)$-form gauge potential $B_{[q+1]}$,
with associated $(q+2)$-form field strength $H_{[q+2]}=\dd B_{[q+1]}$.
The corresponding action is \eqref{theory}.
We want to construct $p$-brane solutions of this theory by uplifting the
Gibbons-Maeda solution. To this end, we start with a metric formed by
the warped product of a $d$-dimensional base space with metric $\tilde
g_\munu$ and coordinates $x^\mu$, and $q$ flat dimensions with
coordinates $y^m$,
\eq
ds^2=e^{2\al\phi(x)}\tilde g_\munu(x) dx^\mu dx^\nu+e^{2\be\phi(x)}\delta_{mn}(y)dy^m dy^n,
\label{metricreductionansatz}\eeq
and an electric ansatz for the gauge potential with the $q$ extra indices lying in the extended directions,
\eq
B_{[q+1]}=\tilde B_{[1]}\wedge
\dd y^{1}\wedge\ldots\wedge\dd y^{p},
\eeq
in such a way that, upon dimensional reduction, the $(q+1)$-form potential manifests itself as a Maxwell potential. We also take the dilaton field to be proportional to its lower dimensional embodiment, $\phi=\ga\tilde\phi$, with $\ga$ a coefficient to be determined. The $D$-dimensional Ricci scalar $R$ of the metric (\ref{metricreductionansatz}) can be written in terms of the Ricci scalar $\widetilde R$ of the base space and the dilaton $\phi$ as,
\eqa
\displaystyle e^{2\al\phi}R=\widetilde R-2\lp(d-1)\al+q\be\rp\Delta\phi\qquad\qquad\qquad\qquad\qquad\qquad\qquad\nonumber\\
\displaystyle
-\lp(d-1)(d-2)\al^2+2q(d-2)\al\be+q(q+1)\be^2\rp(\nabla \phi)^2,
\label{Ricci}\eeqa
Then, after imposing
\eq
(n+1)\al+q\be=0
\label{c0}\eeq
to enforce the Einstein frame, using the relation
\eq
H^2=\frac12(q+2)!\, e^{-4\al\phi-2q\be\phi}\tilde F^2,
\eeq
and integrating by parts a laplacian of the dilaton, the action (\ref{theory}) reduces to
\eq
I=\frac1{16\pi G}\int d^dx\sqrt{-\tilde g}
\lb
\tilde R-\lp\frac{n+q+1}{n+1}q\be+2\rp\ga^2\lp\p\tilde\phi\rp^2
-\frac{ e^{-2\lp a+\al+q\be\rp\ga\tilde\phi}}{4}\tilde F^2
\rb,
\eeq
with $\tilde F=\dd\tilde B_{[1]}$.
This action is the EMD action (\ref{EMDaction}) with dilaton coupling $\tilde a$ given by
\eq
\tilde a =\ga\lp a+\frac{nq}{n+1}\be\rp,
\label{c1}\eeq
as long as one chooses $\ga$ such that the dilaton kinetic term has the correct normalization,
\eq
\lp\frac{n+q+1}{n+1}q\be^2+2\rp\ga^2=2.
\label{c2}\eeq
However this is not sufficient to ensure that solutions of the lower dimensional EMD equations solve also the $D$-dimensional equations of motion. Indeed, varying equation (\ref{theory}) we find
\eq
R_\munu=2\nabla_\mu\phi\nabla_\nu\phi
+\frac{1}{2(q+1)!}e^{-2a\phi}H_{\mu\rho_1\ldots\rho_{q+1}}H_{\nu}{}^{\rho_1\ldots\rho_{q+1}}
-\frac{q+1}{2(D-2)(q+2)!}e^{-2a\phi}H^2g_\munu\,,
\eeq
\eq
\Delta \phi+\frac a{4(q+2)!}e^{-2a\phi}H^2=0,\qquad
\nabla_\nu\lp e^{-2a\phi}H^{\nu\mu_1\ldots\mu_{q+1}}\rp=0.
\eeq
While the field equation for $H$ is automatically satisfied for the uplifted solution, Einstein's equations and the dilaton equantion are not; indeed the equation for the dilaton becomes upon dimensional reduction
\eq
\tilde\Delta\tilde\phi+\frac a{8\ga}e^{-2\tilde a\tilde\phi}\tilde F^2=0,\qquad
\eeq
which is equivalent to the lower dimensional dilaton equation of motion if and only if $a=\gamma\tilde a$.
This relation, with (\ref{c1}), can be solved to fix the value of $\be$ such that both the dilaton equation and the Einstein's equation are verified,
\eq
\be=\frac{2n}{(n+q+1)a}\,.
\eeq
With the latter relation, equations (\ref{c0}), (\ref{c1}) and (\ref{c2}) can be solved for $\al$, $\ga$ and $\tilde a$. In particular, the dilaton couplings are related by
\eq
\tilde a^2=a^2+\frac{2n^2q}{(n+1)(n+q+1)}\,.
\eeq

We can now uplift the Gibbons-Maeda solution to the higher dimensional theory, to generate electrically charged $q$-branes in $n+q+3$ dimensions. The resulting metric is
\eq
ds^2=-\frac1{h^A}\lp f\,dt^2+d\vec y^2\rp+h^B\lp\frac{dr^2}f
+r^2d\Om_{(n+1)}^2\rp,
\label{qbranemetric}\eeq
with
\eq
A=\frac{4n}{2n(q+1)+(n+q+1)a^2}\,,\qquad
B=\frac{4(q+1)}{2n(q+1)+(n+q+1)a^2}\,,
\eeq
the gauge field reads
\eq
B_{[q+1]}=-\sqrt{A+B}\,\frac{r_0^n}{r^nh(r)}\sinh\al\cosh\al\,\dd t\wedge\ep\,,
\eeq
and the dilaton is given by
\eq
\phi=-\frac14\lp A+B\rp a\ln h(r)\,.
\eeq
For future reference, notice that the $A$ and $B$ coefficients enter these formulas through the combination $A+B$, that we dub hereafter $N$, and that they satisfy the relation $(q+1)A-nB=0$.

%%%%%%%%%%%%%%%%%%%%%%%%%%%%%%%%%%%%%%%%%%%%%%%%%%%%%%%%%%%%%%%%%%%%%%%%
\subsection{Black $p$-branes with electric $q$-brane charge}

We can now obtain general $p$-brane solutions with diluted $q$-brane charge ($q\leq p$) by taking the solution of the previous subsection, and uplifting it once more by adding $p-q $ additional extended directions. We keep the gauge field lying in the first $q$ extended directions $y^1$,\ldots, $y^{q}$, and we rename the new extra $(p-q)$ coordinates as $z^a$. We first perfom a dimensional reduction along the $y^m$ directions. Since the procedure is precisely the same as the one explained in the previous subsection, we will give the final solution without any further detail.

The black $p$-brane solution with electric $q$-charge to the theory defined by the action (\ref{theory}) in $D=n+p+3$ dimensions has the metric
\eq
ds^2=-\frac1{h^A}\lp f\,dt^2+d\vec y^2\rp+h^B\lp\frac{dr^2}f+r^2d\Om_{(n+1)}^2+d\vec z^2\rp,
\label{branemetric}\eeq
with
\eq
A=\frac{4(n+p-q)}{2(q+1)(n+p-q)+(n+p+1)a^2}\,,\qquad
B=\frac{4(q+1)}{2(q+1)(n+p-q)+(n+p+1)a^2}\,.
\eeq
Then, defining
\eq\label{Ndef2}
N=A+B=\frac{4(n+p+1)}{2(q+1)(n+p-q)+(n+p+1)a^2}\,,
\eeq
the gauge field reads
\eq
B_{[q+1]}=-\sqrt{N}\,\frac{r_0^n}{r^nh(r)}\sinh\al\cosh\al\,\dd t\wedge\dd y^1\wedge\ldots\wedge\dd y^{q}\,,
\label{Bform}\eeq
and the dilaton field is given by
\eq
\phi=-\frac14N a\,\ln h(r)\,.
\label{dilaton}\eeq

Note that this solution is also valid for non-dilatonic branes. In this case, one simply has to use the previous formulas with $a=0$, giving:
\eq
A=\frac2{q+1}\,,\qquad
B=\frac2{n+p-q}\,,\qquad
N=\frac{2(n+p+1)}{(q+1)(n+p-q)}\,.
\eeq

It is worth mentioning a few other particular cases. When $p=q$, one correctly recovers the solution of the previous subsection. Instead, setting $q=0$ one obtains general $p$-brane solutions of EMD theory,
\eq
A=\frac{4(n+p)}{2(n+p)+(n+p+1)a^2},\quad
B=\frac{4}{2(n+p)+(n+p+1)a^2},\quad
N=\frac{4(n+p+1)}{2(n+p)+(n+p+1)a^2}\,,
\eeq
that reduce by choosing vanishing dilaton coupling $a^{\textsf{EM}}$ to the $p$-branes of pure Einstein-Maxwell theory,
\eq
A^{\textsf{EM}}=2\,,\qquad
B^{\textsf{EM}}=\frac{2}{n+p}\,,\qquad
N^{\textsf{EM}}=2\frac{n+p+1}{n+p}\,,
\eeq
while, by choosing for the dilaton coupling
\eq
a^{\textsf{KK}}=\sqrt{\frac{2(n+p+2)}{n+p+1}}=\sqrt{\frac{2(D-1)}{(D-2)}}
\label{aKK}\eeq
relevant for the Kaluza-Klein theory, one obtains
\eq
A^{\textsf{KK}}=\frac{n+p}{n+p+1},\qquad
B^{\textsf{KK}}=\frac{1}{n+p+1},\qquad
N^{\textsf{KK}}=1\,.
\eeq

Finally, the metric can be written in a manifestly Poincar\'e invariant form to describe boosted branes,
\eq
ds^2=\frac1{h^A}\lp \eta_{ab}+\frac{r_0^n}{r^n}u_au_b+\lp h^N-1\rp\hat\perp_{ab}\rp d\si^a d\si^b
+h^B\lp\frac{dr^2}f+r^2d\Om_{(n+1)}^2\rp,
\label{boostedmetric}\eeq
\eq
B_{[q+1]}=-\sqrt{N}\,\frac{r_0^n}{r^nh(r)}\sinh\al\cosh\al\,u\wedge v^{(1)}\wedge\ldots\wedge v^{(q)}\,,
\label{Bformboosted}\eeq
where $\si^a=(t,\vec y,\vec z)$, $\eta_{ab}$ is the flat Minkowski metric, $\hat\perp_{ab}$ is the projector to the directions orthogonal to the gauge field, and $v^{(1)},\ldots,v^{(q)}$ are an orthonormal basis of the spatial subspace of Minkowski on which the gauge field lies, $\eta_{ab}v^{(i)a}v^{(j)b}=\delta^{ij}$, $\hat\perp_{ab}v^{(i)b}=0$.

%%%%%%%%%%%%%%%%%%%%%%%%%%%%%%%%%%%%%%%%%%%%%%%%%%%%%%%%%%%%%%%%%%%%%%%%%%%%%%
%%%%%%%%%%%%%%%%%%%%%%%%%%%%%%%%%%%%%%%%%%%%%%%%%%%%%%%%%%%%%%%%%%%%%%%%%%%%%%
\section{Charges and thermodynamics of the black branes}
\label{app:charges}

%%%%%%%%%%%%%%%%%%%%%%%%%%%%%%%%%%%%%%%%%%%%%%%%%%%%%%%%%%%%%%%%%%%%%%%%%%%%%%

In this section, we compute the charges and study the thermodynamics of
the black $p$-branes with $q$-brane charges obtained in the previous
section. This will provide the basic input for the blackfold formalism:
the equation of state of the effective fluid describing black holes with
more than one scale. The mass density and the tension of the brane are
obtained from the Brown-York stress tensor at infinity, that we
regularize through the background subtraction technique. The electric
charge density is obtained from the flux at infinity using the Gauss'
law, while the temperature and entropy are derived in the standard way,
resulting in an equation of state and in the thermodynamic relations for an
anisotropic fluid.

Since we are computing worldvolume densities, these are the same
as conserved charges of the original Gibbons-Maeda black hole that these
branes are an uplift of. Thus the results should depend only on the
parameters $n$ and $N$ which are invariant under compactification, and
not on $p$ nor $q$. Our results do indeed verify this.

%%%%%%%%%%%%%%%%%%%%%%%%%%%%%%%%%%%%%%%%%%%%%%%%%%%%%%%%%%%%%%%%%%%%%%%%%%%%%%
\subsection{Brown-York stress tensor: energy density and tension}
\label{app:BY}

The Brown-York quasi-local stress tensor is defined on a boundary
timelike hypersurface, which we take at a fixed large radius $r=R$, and reads
\eq
16\pi G\tau_{\mu\nu}=K_{\mu\nu}-h_{\mu\nu} K - \lp \hat K_{\mu\nu}-h_{\mu\nu} \hat K
\rp,
\eeq
where $K_{\mu\nu}$ is the extrinsic curvature of the boundary and the
hatted variables are the corresponding quantities that we subtract,
computed on flat spacetime with the same intrinsic
geometry on both boundaries.
It is easy to see that for a metric of the form
\eq
ds^2=-\frac f{h^A}\,dt^2+h^B\lp\frac{dr^2}f+r^2d\Om_{(n+1)}^2\rp+h^C d\vec y^2+h^D d\vec z^2,
\eeq
the Brown-York stress tensor reads
\begin{eqnarray*}
&&\tau_{tt}=\frac{1}{16\pi G}\frac{r_0^n}{R^{n+1}}\lp{n+1+\lp qC+(p-q)D+(n+1)B\rp n\sinh^2\al}\rp +\mc{O}\lp\frac1{R^{2n+1}}\rp,\\
&&\tau_{mn}=-\frac{r_0^n}{16\pi G R^{n+1}}\lp1+\lp(q-1)C+(p-q)D+(n+1)B-A\rp n\sinh^2\al\rp\delta_{mn}+\mc{O}\lp\frac1{R^{2n+1}}\rp\de_{mn},\\
&&\tau_{ab}=-\frac{r_0^n}{16\pi G R^{n+1}}\lp1+\lp qC+(p-q-1)D+(n+1)B-A\rp n\sinh^2\al\rp\delta_{ab}+\mc{O}\lp\frac1{R^{2n+1}}\rp\de_{ab},\\
&&\tau_{ij}=-\frac{nr_0^n}{16\pi G R^{n-1}}\lp qC+(p-q)D+nB-A\rp\sinh^2\al\,\si_{ij}+\mc{O}\lp\frac1{R^{2n-1}}\rp\si_{ij},
\end{eqnarray*}
where the indices $m$, $n=1,\ldots,q$ are along the $q$ directions
$y^m$, and the indices $a$, $b=1,\ldots,p-q$ run along the $(p-q)$
directions $z^a$. For completeness we have also included the stress
along the angular
directions $i,j$ of $S^{n+1}$ with $d\Omega_{(n+1)}^2=\sigma_{ij}d\theta^i
d\theta^j$.

We obtain the corresponding blackfold stress tensor by integrating over
the transverse sphere at large $R$, and taking the limit $R\rightarrow\infty$
\cite{Emparan:2009at},
\eqa
&&T_{tt}=\frac{\Omega_{(n+1)}}{16\pi G}r_0^n\lp{n+1+\lp qC+(p-q)D+(n+1)B\rp n\sinh^2\al}\rp,\nonumber\\
&&T_{mn}=-\frac{\Omega_{(n+1)}}{16\pi G}r_0^n\lp1+\lp(q-1)C+(p-q)D+(n+1)B-A\rp n\sinh^2\al\rp\delta_{mn},\qquad \label{branest}\\
&&T_{ab}=-\frac{\Omega_{(n+1)}}{16\pi G}r_0^n\lp1+\lp qC+(p-q-1)D+(n+1)B-A\rp n\sinh^2\al\rp\delta_{ab}\nonumber.
\eeqa
Notice that the integrated tension of the $S^{n+1}$ sphere diverges
unless
\eq
qC+(p-q)D+nB-A=0.
\label{ABCD}\eeq
This is automatically satisfied by all the solutions
obtained in appendix~\ref{app:branesol}, and we will assume it in the
following.

This stress tensor describes an anisotropic perfect fluid in the
$(p+1)$-dimensional worldvolume of the blackfold with energy density
and pressures $P_\parallel$ in the directions of the gauge field and
$P_\perp$ in the orthogonal directions to it along the extended
dimensions given respectively by
\eqa
&&\displaystyle\varepsilon=\frac{\Omega_{(n+1)}}{16\pi G}r_0^n\lp n+1+n N\sinh^2\al\rp\,,
\label{energy}\\
&&P_\parallel=-\frac{\Omega_{(n+1)}}{16\pi G}r_0^n\lp1+\lp B-C\rp n\sinh^2\al\rp,\\
&&P_\perp=-\frac{\Omega_{(n+1)}}{16\pi G}r_0^n\lp1+\lp B-D\rp n\sinh^2\al\rp.
\eeqa
All extensive charges are defined as densities per unit worldvolume of
the blackfold as measured from infinity, \ie we are formally dividing
by the factor $\int\! d^{q}y d^{p-q}z$.
In particular, for all solutions obtained in the previous section, $D=B$
and $C=-A$, and the pressures read
\eq
P_\parallel=-\frac{\Omega_{(n+1)}}{16\pi G}r_0^n\lp1+n N\sinh^2\al\rp,\qquad
P_\perp=-\frac{\Omega_{(n+1)}}{16\pi G}r_0^n.
\eeq
As expected, in its perpendicular directions the gauge field does not
produce any stress, and $P_\perp$ is the tension in the neutral case,
while $P_\parallel$ gets modified by the gauge field stresses. They lead
to the equation of state
\eq
\varepsilon=-P_\parallel-nP_\perp\,.
\eeq

%%%%%%%%%%%%%%%%%%%%%%%%%%%%%%%%%%%%%%%%%%%%%%%%%%%%%%%%%%%%%%%%%%%%%%%%%%%%%%
\subsection{Electric $q$-brane charge and potential}
\label{app:charge}

As usual, we define the electric charge by measuring the electric flux at infinity. To fix the normalization, we will conventionally assume that the electromagnetic current $J^{\mu_1\ldots\mu_{q+1}}$ source due to external matter couples to the gauge field $B_{[q+1]}$ according to the action
\eq
I_{sourced}=I
%\frac1{16\pi G}\int d^Dx\sqrt{-g}\lb R-2\lp\nabla\phi\rp^2-\frac1{2(q+2)!}e^{-2a\phi}H^2\rb
+\frac1{(q+1)!}\int d^Dx\sqrt{-g}\,B_{\mu_1\ldots\mu_{q+1}}J^{\mu_1\ldots\mu_{q+1}}.
\label{Hactionsource}\eeq
where $I$ is \eqref{theory}.
Then, the equations of motion for the gauge field get modified to
\eq
\nabla_\nu\lb e^{-2a\phi}H^{\nu\mu_1\ldots\mu_{q+1}}\rb=-8\pi G J^{\mu_1\ldots\mu_{q+1}}.
\eeq
Using Gauss' law it is then possible to define a conserved electric charge as a measure of the electric flux at infinity,
\eq
\mc Q=\oint{}\ast J=\frac1{8\pi G (q+2)!}\oint e^{-2a\phi}H^{\mu_1\ldots\mu_{q+2}}
dS_{\mu_1\ldots\mu_{q+2}},
\label{Q}\eeq
with the integration measure given by
\eq
dS_{\mu_1\ldots\mu_{q+2}}=(q+2)!\,
v^{(1)}_{[\mu_1}\cdots v^{(q+2)}_{\mu_{k+1}]}\,dV_{(D-q-2)},
\eeq
where $dV_{(D-q-2)}$ is the volume element of the hypersurface through which the flux is measured, and the vectors are the unit normal vectors to this hypersurface. To evaluate it for the charged branes (\ref{branemetric})-(\ref{dilaton}), we integrate over the $\calS^{n+1}\times\R^{p-q}$ hypersuface defined at constant $(t,r,y^1,\ldots,y^{q})$ and take the $r\rightarrow\infty$ limit.
The integration over $\calS^{n+1}$ yields a $\Omega_{(n+1)}$ factor, while we omit the formally infinite $\int dz^{1}\cdots\int dz^{p-q}$ integral to obtain the charge density
\eq
\mc Q=\frac{\Omega_{(n+1)}}{16\pi G}n\sqrt{N}r_0^n\sinh\al\cosh\al\,.
\eeq
It is important here to note that this charge is a density {\em in the orthogonal directions to the gauge field}, {\em i.e.} to obtain the total charge one should not integrate over all $p$ spacelike directions of the brane worldvolume, but only on the $p-q$ directions projected by $\hat\perp_{ab}$ in (\ref{boostedmetric}).
Finally, the electric potential at the horizon, measured with respect to infinity, is defined as
\eq
\Phi=B_{ty^1\cdots y^{q}}(\infty)-B_{ty^1\cdots y^{q}}(r_0)=\sqrt{N}\tanh\al.
\eeq

%%%%%%%%%%%%%%%%%%%%%%%%%%%%%%%%%%%%%%%%%%%%%%%%%%%%%%%%%%%%%%%%%%%%%%%%%%%%%%
\subsection{Anisotropic fluid thermodynamics}\label{app:thermo}

The temperature of these black objects is
\eq
T=\frac n{4\pi r_0} \lp\cosh\al\rp^{-N},
\label{T}\eeq
and the entropy density is obtained from the area density,
\eq
s=\frac{a_H}{4G}=\frac{\Omega_{(n+1)}}{4G}r_0^{n+1}\lp\cosh\al\rp^{N}.
\eeq
Notice that the temperature, entropy density, energy density
(\ref{energy}), charge density
(\ref{Qbrane}) and the electric potential (\ref{phi}) depend only on the
parameters $n$ and $N$ and not on $p$.

With these expressions for the energy, temperature, entropy density,
charge density and electric potential, it is easy to check that the
Smarr relation
\eq
\ep=\frac{n+1}nTs+\Phi\mc Q
\label{Smarr}\eeq
holds, as well as the first law of thermodynamics
\eq
\dd\ep=T\dd s+\Phi\dd\mc Q\,.
\eeq
It is worth mentioning that it is possible to obtain a simple expression for the Gibbs free energy $G(T,\Phi)=\epsilon-Ts-\Phi\mc Q$ in term of its natural variables,
\eq
G(T,\Phi)=\frac{\Omega_{(n+1)}}{16\pi G}
\lp\frac n{4\pi T}\rp^n\left[1-\frac{\Phi^2}{N}\right]^{nN/2},
\label{gibbs}\eeq
from which one can easily verify the relations
\eq
s=-\left.\frac{\p G}{\p T}\right|_{\Phi_H},\qquad
\mc Q=-\left.\frac{\p G}{\p\Phi}\right|_T.
\eeq

For an anisotropic fluid of the kind we discuss we have the Gibbs-Duhem
relations\footnote{Note that $P_\perp$ coincides with $-G$, while $-P_\parallel=\ep-Ts$ is the Helmoltz free energy of the system.}
\eq
\ep+P_\parallel=Ts\,,\qquad \ep+P_\perp=Ts+\Phi\mc Q\,.
\label{GD}\eeq
Together with the
first law as written above, and the equation of state of the fluid,
these equations determine the thermodynamics of the system. The Smarr
relation above is a consequence of them.

From the first law and the Gibbs-Duhem relations we obtain
\eq
dP_\perp=sdT+\mc Qd\Phi\,,\qquad
dP_\parallel=sdT-\Phi d\mc Q\,.
\eeq

%%%%%%%%%%%%%%%%%%%%%%%%%%%%%%%%%%%%%%%%%%%%%%%%%%%%%%%%%%%%%%%%%%%%%%%%%%%%%%
%%%%%%%%%%%%%%%%%%%%%%%%%%%%%%%%%%%%%%%%%%%%%%%%%%%%%%%%%%%%%%%%%%%%%%%%%%%%%%
\section{Comparison with exact black hole solutions}
\label{app:comparison}

%%%%%%%%%%%%%%%%%%%%%%%%%%%%%%%%%%%%%%%%%%%%%%%%%%%%%%%%%%%%%%%%%%%%%%%%%%%%%%
\subsection{Charged rotating MP black holes in KK theory}
\label{app:KKbh}

The general charged rotating solution of the Einstein-Maxwell dilaton theory (\ref{EMDaction}) with dilaton coupling (\ref{aKK}) was generated in \cite{Kunz:2006jd} starting with the $D$-dimensional MP black hole with general angular momenta, uplifting it to pure Einstein-Maxwell theory in $D+1$ dimensions, boosting it along the extra dimension and reducing the resulting solution back to $D$ dimensions, where the boost of the solution becomes an electric charge. Having an exact charged rotating explicit solution for this case will provide us with a setting in which we can test the blackfold approach by comparing the effective theory results with the exact results. The final solution has metric, gauge field and dilaton given by
\eqa
ds^2=\lp1+\frac{mr^{2-\ep}}{\Pi F}\sinh^2\al\rp^{\frac1{D-2}}
\left[-dt^2+\frac{\Pi F dr^2}{\Pi-mr^{2-\ep}}
+\lp r^2+a_i^2\rp\lp d\mu_i^2+\mu_i^2 d\phi_i^2\rp
+\ep r^2d\nu^2
\right.\nonumber\\
\left.\!\!\!\!\!\!\!\!\!
+\frac{mr^{2-\ep}}{\Pi F+mr^{2-\ep}\sinh^2\al}
\lp\cosh\al dt-a_i\mu_i^2 d\phi_i\rp^2
\right],\hphantom{mx}
\eeqa
where repeated $i=1\ldots N$ indices are summed over,
$N=\lfloor\frac{D-1}2\rfloor$ is the number of independent angular
momenta\footnote{We employ this fairly standard notation since there is no
possibility of mistaking this $N$ with \eqref{Ndef}: in this
appendix we are concerned with Kaluza-Klein
solutions for which $N$ in \eqref{Ndef} is 1.} and
$\ep=D-2N-1$ vanishes for odd $D$ and $\ep=1$ for even $D$.
The direction cosines $\mu_i$ verify the constraint
\eq
\sum_{i=1}^N\mu_i^2+\ep\nu^2=1.
\eeq
so that the coordinates $(\mu_i,\phi_i)$ describe the $N$ independent
planes of the $D-2$ transverse sphere. The functions $F$ and $\Pi$ read
respectively
\eq
F=1-\sum_{i=1}^N\frac{a_i^2\mu_i^2}{r^2+a_i^2}\,,\qquad \Pi = \prod_{i=1}^N(r^2+a_i^2)\,.
\eeq
Finally, the gauge field and the dilaton assume the form,
\eq
A=-\frac{mr^{2-\ep}\sinh\al}{\Pi F+mr^{2-\ep}\sinh^2\al}\lp\cosh\al\,\dd t-a_i\mu_i^2\,\dd\phi_i\rp
\eeq
\eq
\phi=-\frac14 a^{\textsf{KK}}\ln\lp 1+\frac{mr^{2-\ep}}{\Pi F}\sinh^2\al\rp
\eeq

The ultraspinning limit for this solution can be obtained in the usual
way, we send $s$ rotation parameters $a_j$ with $j=1,\ldots,s$ to
infinity, keeping the $a_k$ with $k=s+1,\ldots,N$ finite, zooming into
the region close to the poles by keeping the new coordinates
$\sigma_j=a_j\mu_j$ finite in the limiting process, and rescaling the
mass parameter such that $\hat m$, defined by
\eq
\frac m{\prod a_j^2}=\hat m\,,
\eeq
remains finite. Then the $(\si_j,\phi_j)$ coordinates describe $s$ (conformally) flat planes, and we use $p=2s$ cartesian coordinates $\vec y$ to parameterize the corresponding $\R^p$.
The resulting field configuration is
\eqa
ds^2=\lp1+\frac{\hat mr^{2-\ep}}{\hat \Pi\hat F}\sinh^2\al\rp^{\frac1{D-2}}
\left[-dt^2+\frac{\hat\Pi\hat F dr^2}{\hat\Pi-\hat mr^{2-\ep}}
+\lp r^2+a_k^2\rp\lp d\mu_k^2+\mu_k^2 d\phi_k^2\rp
+\ep r^2d\nu^2
\right.\nonumber\\
\left.
+\frac{\hat mr^{2-\ep}}{\hat\Pi\hat F+\hat mr^{2-\ep}\sinh^2\al}
\lp\cosh\al dt-a_k\mu_k^2 d\phi_k\rp^2+d\vec y^2
\right]\hphantom{mm}
\eeqa
where $k$ indices now range from $s+1$ to $N$ and the remaining direction cosines $\mu_k$ verify the constraint
\eq
\sum_{k=s+1}^N\mu_k^2+\ep\nu^2=1.
\eeq
so that the coordinates $(\mu_i,\phi_i)$ describe the $N$ independent planes of the $D-2$ transverse sphere.
The limiting functions $\hat F$ and $\hat\Pi$ read instead
\eq
\hat F=1-\sum_{k=s+1}^N\frac{a_k^2\mu_k^2}{r^2+a_k^2}\,,\qquad \hat\Pi = \prod_{k=s+1}^N(r^2+a_k^2)\,.
\eeq
and the gauge and dilaton fields assume the form,
\eq
A=-\frac{\hat mr^{2-\ep}\sinh\al}{\hat\Pi\hat F+\hat mr^{2-\ep}\sinh^2\al}\lp\cosh\al\,\dd t-a_k\mu_k^2\,\dd\phi_k\rp
\eeq
\eq
\phi=-\frac14 a^{\textsf{KK}}\ln\lp 1+\frac{\hat mr^{2-\ep}}{\hat\Pi\hat F}\sinh^2\al\rp\,.
\eeq
These solutions represent black membranes, with $p$ extended dimensions, whose horizon is $\R^p\times\calS^{n+1}$, and with arbitrary angular momenta on the $(n+1)$-sphere ($n$ is defined as $n=D-p-3$).

As we shall not consider in this work the angular momentum on the sphere, we set $a_k=0$ and we obtain, setting $\hat m=r_0^n$,
\eq
ds^2=-\frac f{h^{\frac{n+p}{n+p+1}}}\,dt^2+h^{\frac1{n+p+1}}\lp\frac{dr^2}f+d\vec y^2+r^2 d\Omega_{n+1}^2\rp
\label{kkbrane}\eeq
with
\eq
A=-\frac{r_0^n\sinh2\al}{2r^nh(r)}\,\dd t\,,\qquad
\phi=-\frac14 a^{\textsf{KK}}\ln h(r)\,.
\eeq

The physical properties of the charged MP dilaton black hole are
\eq
M=\frac{\Om_{(D-2)}}{16\pi G}m\lp1+(D-3)\cosh^2\al\rp\,,\qquad
Q=\frac{(D-3)\Om_{(D-2)}}{16\pi G}m\sinh\al\cosh\al\,,
\eeq
\eq
J_i=\frac{\Om_{(D-2)}}{8\pi G}ma_i\cosh\al\,,\qquad
\Om_i=\frac{a_i}{\lp r_+^2+a_i^2\rp\cosh\al},
\eeq
\eq
\mathcal{A}_{\mathrm h}=\Om_{(D-2)}\frac{\cosh\al}{r_+^{1-\ep}}
\prod_{i=1}^N\lp r_+^2+a_i^2\rp\,,\qquad
\ka=\frac1{\cosh\al}\lp\sum_{i=1}^N\frac{r_+}{r_+^2+a_i^2}-\frac{2-\ep}{2r_+}\rp,
\eeq
where $r_{+}$ is the largest non-negative root of $\prod_{i} (r^{2}+a_{i}^{2})-mr^{2-\epsilon}=0$.\\

In the ultraspinning limit, with $s$ ultraspins such that  $a_i\gg m^{1/(D-3)}$, we obtain
 \eq
M\rightarrow\frac{\Om_{(D-2)}}{16\pi G}r_+^n\lp\prod_{i=1}^s a_i^2\rp\lp1+(D-3)\cosh^2\al\rp\,,\quad
Q\rightarrow\frac{(D-3)\Om_{(D-2)}}{16\pi G}r_+^n\lp\prod_{i=1}^sa_i^2\rp\sinh\al\cosh\al\,,
\eeq
\eq
J_i\rightarrow\frac{\Om_{(D-2)}}{8\pi G}r_+^na_i\cosh\al\prod_{j=1}^sa_j^2\,,\qquad
\Om_i\rightarrow\frac{1}{a_i\cosh\al}
\eeq
\eq
\mathcal{A}_{\mathrm h}\rightarrow\Om_{(D-2)}r_+^{n+1}\cosh\al\prod_{i=1}^sa_i^2
\,,\qquad
\ka\rightarrow\frac n{2r_+\cosh\al}\,,
\eeq
 where we have defined
 \eq
 n=D-2s-3\,.
 \eeq
Then, eliminating $r_+$ and $a_i$ in favor of the temperature $T=\ka/2\pi$ and $\Om_i$, we obtain in the ultraspinning regime,
\eqa
&&M\rightarrow\frac{\Om_{(D-2)}}{16\pi G}\lp\frac n{4\pi T}\rp^n\frac{1+(D-3)\cosh^2\al}{(\prod\Om_i^2)\cosh^{D-3}\al},\quad
Q\rightarrow\frac{(D-3)\Om_{(D-2)}}{16\pi G}\lp\frac n{4\pi T}\rp^n\frac{\sinh\al\cosh\al}{(\prod\Om_i^2)\cosh^{D-3}\al},\nonumber\\
&&J_i\rightarrow\frac{\Om_{(D-2)}}{8\pi G}\lp\frac n{4\pi T}\rp^n\frac{\cosh^{-(D-3)}\al}{\Om_i\prod_j\Om_j^2},\qquad\qquad
\mathcal{A}_{\mathrm h}\rightarrow\Om_{(D-2)}\lp\frac n{4\pi T}\rp^{n+1}\frac{\cosh^{-(D-3)}\al}{\prod_i\Om_i^2}.
\eeqa
We can compare these results in the single spinning case with the blackfold results obtained in section \ref{sectiondisk0}. Taking $s=1$, and using the relation
\eq
\Om_{(D-2)}=\Om_{(n+3)}=\frac{2\pi}{n+2}\Om_{(n+1)}
\eeq
the ultraspinning charges reduce to
\eqa
&&M=\frac{\Omega_{(n+1)}}{8G(n+2)\Om^2}\lp\frac n{4\pi T}\rp^n
\frac{1+(n+2)\cosh^2\al}{\lp\cosh\al\rp^{n+2}},\quad
J=\frac{\Omega_{(n+1)}}{4G(n+2)\Om^3}\lp\frac n{4\pi T}\rp^n
\lp\cosh\al\rp^{-(n+2)},\nonumber\\
&&{\mathcal A}_{\mathrm h}=\frac{2\pi\Omega_{(n+1)}}{(n+2)\Om^2}\lp\frac n{4\pi T}\rp^{n+1}
\lp\cosh\al\rp^{-(n+2)},\qquad\quad
Q=\frac{\Omega_{(n+1)}}{8G\Om^2}\lp\frac n{4\pi T}\rp^n
\frac{\sinh\al\cosh\al}{\lp\cosh\al\rp^{n+2}}.
\eeqa
Substituting the parameter $\al$ for the electric potential defined by
$\Phi_H=\tanh\al$ and the horizon area for $S={\mc A}_{\mathrm h}/4G$, it can readily be checked these
quantities, describing the ultraspinning regime of the MP dilaton black
hole, coincide with the charges \eqref{diskKK} obtained using the
blackfold approach.

The ultimate meaning of the agreement between our blackfold construction
of KK-charged ultraspinning black holes, and the ultraspinning limit of
the exact rotating KK solutions, is that the procedure of adding KK
charge by boosting in the KK direction commutes with the ultraspinning
limit.

\subsection{Charged black rings in five dimensions}
\label{app:chring}

The solution we consider is a particular case of those derived in
\cite{Elvang:2003mj,Elvang:2004xi}, and we use the notation of the
latter reference. We set to zero two of the charge parameters $\alpha_i$
and all three independent dipole parameters $\mu_i$. The solution will
then have a charge and only the dipole that is induced by the rotation
of the charge. Otherwise, the blackfold that would describe it should
have both 0-brane and 1-brane charge, which is possible but is not
addressed in this paper.

The solution is
\beqa
ds^2&=&
-h^{-2/3}\frac{F(y)}{F(x)}\lp dt+R(1+y)\frac{C_\la}{F(y)}\cosh\al d\psi\rp^2\nonumber\\
&&
+h^{1/3}F(x)\frac{R^2}{(x-y)^2}\lp -\frac{G(y)}{F(y)}d\psi^2-\frac{dy^2}{G(y)}+\frac{dx^2}{G(x)}
+\frac{G(x)}{F(x)}d\phi^2\rp\,,
\eeqa
\beqa
A&=&\frac{\la (y-x)}{h F(x)}\cosh\al \sinh\al\, dt
+R C_\la \frac{1+y}{h F(x)}\sinh\al\, d\psi
\,,\\
\phi&=&-\frac1{\sqrt{6}}\ln h\,,
\eeqa
where
\beq
F(\xi)=1+\la\xi\,,\qquad G(\xi)=(1-\xi^2)(1+\nu\xi)\,,\qquad
h=1+\frac{\la(x-y)}{F(x)}\sinh^2\al\,,
\eeq
\beq
C_\la =\sqrt{\la(\la-\nu)\frac{1+\la}{1-\la}}\,,
\eeq
and the equilibrium of the ring requires
\beq\label{ringequil}
\la=\frac{2\nu}{1+\nu^2}\,.
\eeq

The ultraspinning, infinite radius limit of this solution is obtained by
taking $R\to\infty$, $\lambda,\nu\to 0$ while keeping fixed $\alpha$ and
\beq
R/y=-r,\qquad \nu R=r_0\,,\qquad \lambda
R=r_0(\cosh^2\eta+\sinh^2\eta\sinh^2\alpha)\,
\eeq
(the apparent difference in the scaling of $\lambda$ relative to
\cite{Elvang:2004xi} is due to the different order in which the momenta
in the KK direction and in the string direction are taken, \ie the
boosts are measured in different frames). In this
limit the solution becomes the same as the boosted string with KK charge
in \eqref{boostedmetric} with $A=2/3$, $B=1/3$, $N=1$. The equilibrium
condition \eqref{ringequil} now becomes
\beq
\cosh^2\eta+\sinh^2\eta\sinh^2\alpha=2\,,
\eeq
\ie
\eq
\sinh^2\eta\cosh^2\alpha=1\,,
\eeq
which is exactly the same as \eqref{equilsph} with $p=n=N=1$. Since in
the limit $R\to\infty$ it has been shown that the mass, angular
momentum, charge and entropy of the black ring are equal to the
integrals of worldsheet densities \cite{Elvang:2003mj}, this agreement
between the equilibrium boosts is sufficient to guarantee that the
blackfold construction reproduces correctly all physical properties of
the ultraspinning black ring.

We might also consider two non-zero charges but this does not add
anything qualitatively new. The non-dilatonic solutions $N=3$ in
\cite{Elvang:2004xi} appear to fall outside the scope of our
construction due to the effect of the Chern-Simons term in the action.

%%%%%%%%%%%%%%%%%%%%%%%%%%%%%%%%%%%%%%%%%%%%%%%%%%%%%%%%%%%%%%%%%%%%%%%%
\subsection{Dipole black rings in five dimensions}
\label{app:dipring}

Ref.~\cite{Emparan:2004wy} presented exact solutions for
five-dimensional black rings with string dipole. It was also shown that
in the ultraspinning regime the solution limits to a boosted black
string with string charge, and with a specific value of the boost. This
value is exactly the same that we have obtained in \eqref{equileta}, for
$n=1$, $p=1$, and any $N$.
%Ref.~\cite{Emparan:2004wy} had already
%checked that the properties of the dipole black ring in its
%ultraspinning limit could be obtained from those of a boosted black
%string with this value of the boost (in what was essentially the
%blackfold calculation of physical quantities).
For the same reason as in the previous example of charged rings, it
follows that our blackfold
construction correctly reproduces the physics of five-dimensional dipole
black rings to leading order in the ultraspinning regime.

%%%%%%%%%%%%%%%%%%%%%%%%%%%%%%%%%%%%%%%%%%%%%%%%%%%%%%%%%%%%%%%%%%%%%%%%%%%%%%
%%%%%%%%%%%%%%%%%%%%%%%%%%%%%%%%%%%%%%%%%%%%%%%%%%%%%%%%%%%%%%%%%%%%%%%%%%%%%%

%%%%%%%%%%%%%%%%%%%%%%%%%%%%%%%%%%%%%%%%%%%%%%%%%%%%%%%%%%%%%%%%%%%

\end{document}